\documentclass[aps,prd,floats,superscriptaddress,floatfix,nofootinbib,preprintnumbers,notitlepage,eqsecnum]{revtex4-1}
\usepackage{amsmath}
\usepackage{amsfonts}
\usepackage{graphicx}
\usepackage{slashed}
\usepackage{dcolumn}
\usepackage{bm}
\usepackage{amssymb}
\usepackage{latexsym}
\usepackage{color}
\usepackage{url}

\newcommand{\pbrac}[1]{\left( #1 \right)}
\newcommand{\tbrac}[1]{\left[ #1 \right]}
\newcommand{\cbrac}[1]{\left\{ #1 \right\}}

\begin{document}

\title{Prospects for Discovering the Higgs-like Pseudo-Nambu-Goldstone Boson\\of the Classical Scale Symmetry}

\author{Arsham Farzinnia}
\email[]{farzinnia@ibs.re.kr}
\affiliation{Center for Theoretical Physics of the Universe, Institute for Basic Science (IBS)\\Daejeon 305-811, Republic of Korea}

\preprint{CTPU-15-10}

\date{\today}

\begin{abstract}
We examine the impact of the expected reach of the LHC and the XENON1T experiments on the parameter space of the minimal classically scale invariant extension of the standard model (SM), where all the mass scales are induced dynamically by means of the Coleman-Weinberg mechanism. In this framework, the SM content is enlarged by the addition of one complex gauge singlet scalar with a scale invariant and $CP$-symmetric potential. The massive pseudoscalar component, protected by the $CP$-symmetry, forms a viable dark matter candidate, and three flavors of the right-handed Majorana neutrinos are included to account for the nonzero masses of the SM neutrinos via the see-saw mechanism. The projected constraints on the parameter space arise by applying the ATLAS heavy Higgs discovery prospects, with an integrated luminosity of 300 and 3000~fb$^{-1}$ at $\sqrt s = 14$~TeV, to the pseudo-Nambu-Goldstone boson of the (approximate) scale symmetry, as well as by utilizing the expected reach of the XENON1T direct detection experiment for the discovery of the pseudoscalar dark matter candidate. A null-signal discovery by these future experiments implies that vast regions of the model's parameter space can be thoroughly explored; the combined projections are expected to confine a mixing between the SM and the singlet sector to very small values, while probing the viability of the TeV~scale pseudoscalar's thermal relic abundance as the dominant dark matter component in the Universe. Furthermore, the vacuum stability and triviality requirements of the framework up to the Planck scale are studied and the viable region of the parameter space is identified. The results are summarized in extensive exclusion plots, incorporating additionally the prior theoretical and experimental bounds for comparison.
\end{abstract}

\maketitle

\section{Introduction}\label{intro}

Classical scale invariance has been proposed as an interesting theoretical route to connect various mass scales present within a particular formal framework in a technically natural manner \cite{natural}. Within a classically scale invariant theory, the quadratic sensitivity of one mass scale to another is always accompanied by a mutual coupling, which is not a free parameter, but the magnitude of which is set by the scale symmetry to be proportional to the ratio of the two mass scales. As a consequence, any particular mass scale is naturally protected from a quadratic destabilization due to any other scale within the same framework.\footnote{See \cite{smallcouple} for interesting discussions in this regard.} This is attributed to the fact that the hard logarithmic breaking of the classical scale invariance, by means of the dimension-4 operators at the quantum level, cannot affect the dimension-2 mass operators \cite{Bardeen}.\footnote{At this point, one should emphasize that such a protection from a mutual quadratic destabilization is only granted for the mass scales incorporated \textit{within} the same scale symmetric framework. These mass scales will still be quadratically sensitive to any other higher mass threshold \textit{outside} of the scale symmetric framework with an $\mathcal O(1)$~coupling.}

As an attractive application, one may consider constructing a classically scale invariant version of the standard model (SM), by removing the mass term of the SM Higgs field $\mu_H^{2}$---constituting the sole dimensionful parameter of the SM and responsible for the spontaneous breaking of the electroweak symmetry---from the classical potential. It is well-known that the quantum corrections lead to the radiative generation of such nonzero Higgs field mass term by means of the dimensional transmutation \cite{Coleman:1973jx}; nevertheless, the obtained (loop-induced) Higgs boson mass, identified as the pseudo-Nambu-Goldstone boson of the (approximate) scale symmetry, is parametrically too light to conform with the LEP-II limit ($M_{h} \gtrsim 114$~GeV) \cite{LEPII}, signifying the failure of the Coleman-Weinberg mechanism within the pure SM.

As demonstrated in \cite{Farzinnia:2013pga}, in light of the mentioned failure, it is possible to extend the SM content minimally by one \textit{complex} gauge singlet scalar,\footnote{As discussed in \cite{Farzinnia:2013pga}, the addition of only one \textit{real} singlet scalar to the SM is still inadequate to generate a $M_{h} = 125$~GeV Higgs boson mass, while conforming with the theoretical and experimental constraints. Minimal singlet-extensions of the classically scale invariant SM have also been considered in \cite{SIother,Gabrielli:2013hma,AlexanderNunneley:2010nw}.} in order to accommodate the classical scale symmetry, a successful spontaneous breaking of the electroweak symmetry, as well as a Higgs boson mass $M_{h} = 125$~GeV as discovered by the LHC \cite{LHCnew}. The radiative generation of a nonzero singlet vacuum expectation value (VEV) is transmitted to the Higgs sector by means of the ``Higgs portal'' operators within the potential, inducing a nonzero VEV for the SM Higgs boson, and resulting in a successful spontaneous electroweak symmetry breaking. In the scalar mass basis, the mass of the (mostly SM-like) physical Higgs boson is generated in the classical potential by the dynamically-induced VEVs, whereas the other (mostly singlet-like) $CP$-even scalar obtains a radiative mass at the quantum level. The latter, thus, assumes the role of the pseudo-Nambu-Goldstone boson of the (approximate) scale symmetry within the scenario. Furthermore, demanding the extended scalar potential to be $CP$-invariant renders the $CP$-odd pseudoscalar boson stable, and a suitable WIMP dark matter candidate. The framework additionally accommodates the generation of nonzero masses for the SM neutrinos, utilizing the see-saw mechanism \cite{seesaw}, by introducing three flavors of the right-handed Majorana neutrinos. The latter's masses are, subsequently, generated via their $CP$- and scale invariant Yukawa interactions with the complex singlet.

The LEP \cite{LEPII} and LHC \cite{LHCHeavyH} collider searches for the singlet-like $CP$-even boson, as well as the results from the LUX direct detection experiment \cite{LUX2013} and the (updated) Planck thermal relic abundance \cite{Ade:2015xua} applied to the pseudoscalar WIMP, have been extensively explored in \cite{Farzinnia:2014xia}, identifying the viable region of the model's parameter space. The possibility of inducing a strongly first-order electroweak phase transition within the scenario, relevant to the baryogenesis paradigm, has been investigated in \cite{Farzinnia:2014yqa}. In the current treatment, we consider the expected impact on the parameter space by the discovery prospects of the future LHC and the XENON1T direct detection experiments. Specifically, we use the ATLAS projections for discovering a heavy SM-like Higgs within the mass range 200-1000~GeV \cite{ATLASHeavyS}, with an integrated luminosity of 300 and 3000~fb$^{-1}$ at $\sqrt s = 14$~TeV,\footnote{While the ATLAS Collaboration reports the discovery prospects for a heavy SM-like Higgs boson, the CMS projections pertain to a heavy Higgs within the context of the two-Higgs-doublets model \cite{CMSHeavyS}. In the current treatment, we use the ATLAS SM-like projections.} and the XENON1T expected reach for discovering a WIMP dark matter \cite{Aprile:2012zx} to determine the expected bounds on the parameter space in the event of no signal discovery. In addition, we extend the previous analysis of the vacuum stability and triviality bounds of the scenario performed in \cite{Farzinnia:2013pga}, by consistently increasing the cutoff from 100~TeV to the Planck scale, and determine the viable region of the parameter space.\footnote{The vacuum stability and triviality conditions for our introduced potential \eqref{V0}, without the inclusion of the right-handed Majorana neutrinos, were also investigated in \cite{Gabrielli:2013hma}, with the cutoff set near the energy scale of the known SM $U(1)_{Y}$~Landau pole.}

The treatment is organized as follows: Sec.~\ref{rev} is devoted to reviewing the basic formal ingredients of the minimal classically scale invariant extension of the SM and to define the notation, closely following \cite{Farzinnia:2013pga}. In Sec.~\ref{prosp}, we explore the ATLAS and XENON1T discovery prospects and their impact on the parameter space of the framework if no heavy Higgs or WIMP dark matter is discovered by these experiments. The results are exhibited in various exclusion plots, covering the relevant parameter space, and incorporating the previously studied theoretical and experimental bounds for comparison. We proceed, in Sec.~\ref{TrivStab}, to investigate the regions of the parameter space allowing for the vacuum stability and triviality conditions to be satisfied up the to the Planck scale, discussing the findings according to various exclusion plots. Finally, we conclude by providing a summary of the presented work in Sec.~\ref{concl}.

\section{Review of the Minimal Model}\label{rev}

We start by reviewing the formalism of the minimal classically scale invariant extension of the SM \cite{Farzinnia:2013pga}. In this setup, the mass term of the SM Higgs field, $\mu_H^{2}$, responsible for the spontaneous breaking of the electroweak symmetry, is absent within the classical potential, as required by the scale symmetry. The SM scalar sector is, subsequently, minimally extended by one complex gauge singlet scalar, with the potential containing solely the dimension-4 marginally-relevant operators without any explicit mass scale, and additionally respecting the $CP$-symmetry. In addition, three flavors of the right-handed Majorana neutrinos (with Yukawa couplings to the complex singlet) are included to account for the nonzero masses of the SM neutrinos, by means of the see-saw mechanism \cite{seesaw}.

\subsection{Tree-level Scale- and $CP$-symmetric Scalar Potential}

The most general complex singlet-extended scalar potential, invariant under the scale- and the $CP$-symmetry, is of the form
\begin{equation}\label{V0}
V^{(0)}(H,S) = \frac{\lambda_1}{6} \pbrac{H^\dagger H}^2 + \frac{\lambda_2}{6} |S|^4 + \lambda_3 \pbrac{H^\dagger H}|S|^2 + \frac{\lambda_4}{2} \pbrac{H^\dagger H}\pbrac{S^2 + S^{*2}} + \frac{\lambda_5}{12} \pbrac{S^2 + S^{*2}} |S|^2 + \frac{\lambda_6}{12} \pbrac{S^4 + S^{*4}} \ ,
\end{equation}
where, $H$ and $S$ denote the SM Higgs field and the complex scalar, respectively,
\begin{equation}\label{HS}
H= \frac{1}{\sqrt{2}}
\begin{pmatrix} \sqrt{2}\,\pi^+ \\ v_\phi+\phi+i\pi^0 \end{pmatrix} \ , \qquad S =\frac{1}{\sqrt 2} \pbrac{v_\eta + \eta + i\chi} \ .
\end{equation}
In \eqref{HS}, the $\pi^{0, \pm}$ represent the electroweak Nambu-Goldstone bosons eaten by the $Z, W^{\pm}$ vector bosons, and the $CP$-odd pseudoscalar component, $\chi$, is the WIMP dark matter candidate \cite{Farzinnia:2014xia}. The $CP$-even scalars, $\phi$ and $\eta$, obtain their nonzero VEVs, $v_{\phi}( = 246$~GeV) and $v_{\eta}$, dynamically via the Coleman-Weinberg mechanism \cite{Coleman:1973jx}.\footnote{In principle, the $CP$-odd pseudoscalar component of the singlet may also obtain a nonzero VEV, $v_{\chi}$, leading to a spontaneous violation of the $CP$-symmetry \cite{AlexanderNunneley:2010nw}. For the purposes of the current analysis, however, we assume that this is not the case.}

In addition to the $H$ and $S$~fields' quartic self-interactions, parametrized by the couplings $\lambda_{1}$, $\lambda_{2}$, $\lambda_{5}$ and $\lambda_{6}$, the scalar potential \eqref{V0} offers two ``Higgs portal'' terms, parametrized by $\lambda_{3}$ and $\lambda_{4}$, representing the mutual interaction between the $H$ and $S$~fields. In particular, one observes that once the $CP$-even component of the $S$~field acquires a dynamically-generated nonzero VEV, $v_{\eta}$, a nonzero mass term for the SM Higgs field, $H$, is dynamically induced and identified as
\begin{equation}\label{muSM}
\mu_H^{2} = \frac{\lambda_{3}+\lambda_{4}}{2}\, v_{\eta}^{2} \equiv \frac{\lambda_{m}^{+}}{2}\, v_{\eta}^{2} \ , 
\end{equation}
where, we have defined $\lambda_{m}^{+} \equiv \lambda_{3} + \lambda_{4}$ for later convenience (c.f. \eqref{couprel}). As we shall demonstrate momentarily, the mixing coupling $\lambda_{m}^{+}<0$, giving rise to a successful spontaneous electroweak symmetry breaking.

The potential \eqref{V0} is bounded from below for the following relations between the scalar couplings \cite{Farzinnia:2014xia}
\begin{equation}\label{stabtree}
\begin{split}
&\lambda_\phi^{} > 0 \ , \qquad \lambda_\eta^{} > 0 \ , \qquad
\lambda_\chi^{} > 0 \,, \qquad
\lambda_{\eta\chi}^{} \geq -\frac{1}{3}\!\sqrt{\lambda_{\eta}^{}\lambda_{\chi}^{}} \ , \qquad
\lambda_m^+ \geq -\frac{1}{3} \sqrt{\lambda_\phi \lambda_\eta} \ , \qquad
\lambda_m^- \geq -\frac{1}{3} \sqrt{\lambda_\phi \lambda_\chi} \ , \\
&\lambda_{\eta\chi} \sqrt {\lambda_\phi} + \lambda_m^+ \sqrt{\lambda_{\chi}} + \lambda_m^- \sqrt{\lambda_{\eta}} \geq -\frac{1}{3} \tbrac{\sqrt{\lambda_{\phi} \lambda_{\eta} \lambda_{\chi}} + \sqrt{2\pbrac{3\lambda_{\eta\chi}+\sqrt{\lambda_{\eta} \lambda_{\chi}}}\pbrac{3\lambda_{m}^{+}+\sqrt{\lambda_{\phi} \lambda_{\eta}}}\pbrac{3\lambda_{m}^{-}+\sqrt{\lambda_{\phi} \lambda_{\chi}}}}} \ ,
\end{split}
\end{equation}
with the quartic interactions of the components defined in terms of the original couplings in the potential according to
\begin{equation}\label{couprel}
\lambda_\phi \equiv \lambda_{1} \ , \quad \lambda_\eta \equiv \lambda_{2} + \lambda_{5} + \lambda_{6} \ , \quad \lambda_\chi \equiv \lambda_{2} - \lambda_{5} + \lambda_{6} \ , \quad \lambda_{\eta \chi} \equiv \frac{1}{3}\lambda_{2} - \lambda_{6} \ , \quad \lambda_m^\pm \equiv \lambda_{3} \pm \lambda_{4} \ .
\end{equation}
In terms of these definitions, the quartic interactions between the scalar components can then be conveniently expressed as 
\begin{equation}\label{V0quart}
\begin{split}
V^{(0)}_{\text{quartic}} =&\, \frac{1}{24} \tbrac{ \lambda_\phi \phi^4 + \lambda_\eta \eta^4 + \lambda_\chi \chi^4 + \lambda_\phi\pbrac{\pi^0\pi^0 + 2 \pi^+ \pi^-}^2}+ \frac{1}{4}\tbrac{\lambda_m^+ \phi^2 \eta^2 + \lambda_m^- \phi^2 \chi^2 +\lambda_{\eta \chi} \eta^2 \chi^2}
\\
&+ \frac{1}{12} \tbrac{\lambda_{\phi} \phi^2 + 3\lambda_m^+ \eta^2 + 3\lambda_m^- \chi^2}\pbrac{\pi^0\pi^0 + 2 \pi^+ \pi^-}\ .
\end{split}
\end{equation}

As the nonzero scalar VEVs are induced dynamically at the quantum level, one needs, in principle, to calculate and minimize the full one-loop potential, in order to determine the physical vacuum. This is, however, a formidable endeavor, which might not be possible to perform analytically. Fortunately, the minimization of the potential may be achieved perturbatively, according to the procedure developed by Gildener and Weinberg  \cite{Gildener:1976ih}. To this end, one initially minimizes the tree-level potential \eqref{V0} with respect to its constituent fields $H$ and $S$,
\begin{equation}\label{pertmin}
\frac{d V^{(0)}}{d H} \Big |_{\phi = v_{\phi}} = \frac{d V^{(0)}}{d S} \Big |_{\eta = v_{\eta}} = 0 \ .
\end{equation}
Since in the full quantum theory the couplings run as a function of the renormalization scale, the tree-level perturbative minimization \eqref{pertmin} occurs at a definite energy scale, $\Lambda$. This tree-level minimization defines, subsequently, a flat direction between the fields where the $CP$-even scalars' field values are related to the quartic couplings according to
\begin{equation}\label{flatdir}
\frac{v_\phi^2}{v_\eta^2} = \frac{-3\lambda_m^+(\Lambda)}{\lambda_\phi(\Lambda)}=\frac{\lambda_\eta(\Lambda)}{-3\lambda_m^+(\Lambda)} \ .
\end{equation}
Thus, one needs to consider the quantum corrections only along this flat direction, where the tree-level potential is minimized and the one-loop contributions to the potential are dominant. Once the one-loop potential is taken into account along this flat direction, the tree-level flatness is lifted and the true vacuum of the system is determined.

The Higgs portal parameter, $\lambda_{m}^{+}$, gives rise to a mixing between the $CP$-even scalars, $\phi$ and $\eta$, with dynamically-induced nonzero VEVs. These scalars may be orthogonally rotated, with a mixing angle $\omega$, into the physical mass basis, according to~\cite{Farzinnia:2013pga}
\begin{equation}\label{hs}
\begin{pmatrix} \phi\\ \eta \end{pmatrix}
= \begin{pmatrix} \cos\omega & \sin\omega \\ -\sin\omega & \cos\omega \end{pmatrix} \begin{pmatrix} h \\ \sigma \end{pmatrix} \ , \qquad \tan\omega = \frac{v_{\phi}}{v_{\eta}}\ ,
\end{equation}
along the flat direction of the potential, where the conditions \eqref{flatdir} hold. The $h$ and $\sigma$~fields are then the physical $CP$-even scalars of the model. The dynamically-induced nonzero VEVs generate the following tree-level mass terms for the bosons along the flat direction\footnote{See \cite{Farzinnia:2013pga} for the general treatment  and determination of the bosonic mass terms, valid away from the flat direction.}
\begin{equation}\label{masstreemincond}
M_h^2 = \frac{v_\phi^2}{3 \cos^{2}\omega} \, \lambda_\phi(\Lambda)\ , \qquad M_\chi^2 =  \frac{v_\phi^2}{2} \tbrac{\lambda_m^-(\Lambda) + \lambda_{\eta\chi}(\Lambda) \cot^{2}\omega} \ , \qquad M_\sigma^2 =M_{\pi^0}^2 = M_{\pi^\pm}^2 = 0 \ ,
\end{equation}
where $v_{\eta}$ is eliminated in favor of the mixing angle, $\omega$, using \eqref{hs}. Identifying the $h$~boson with the scalar discovered at the LHC \cite{LHCnew}, one has $M_{h} = 125$~GeV, while the electroweak Nambu-Goldstone bosons, $\pi^{0,\pm}$ remain massless to all orders in perturbation theory. The $\sigma$~scalar, serving as the pseudo-Nambu-Goldstone boson of the (approximate) scale symmetry, is on the other hand only massless at tree-level, and obtains a radiative mass due to the one-loop potential, as discussed in the following subsection.

Interestingly, within the current classically scale invariant framework, one may also incorporate the see-saw mechanism \cite{seesaw} to account for the experimentally observed nonzero masses of the SM neutrinos. To this end, three massive flavors of the gauge-singlet right-handed Majorana neutrinos, $\mathcal N^{i}$, are introduced, which obtain their masses via Yukawa couplings with the complex singlet
\begin{equation}\label{LRHN}
\mathcal{L}_{\mathcal N} = \text{kin.} - \tbrac{Y^\nu_{ij}\, \bar{L}_{\ell}^{i} \tilde{H} \mathcal{N}^{j} + \text{h.c.}} -\frac{1}{2}y_{N} \pbrac{S + S^*} \bar{\mathcal{N}}^{i}\mathcal{N}^{i} \ ,
\end{equation}
with $\mathcal{N}_{i} = \mathcal{N}_{i}^{c}$ the 4-component right-handed Majorana neutrino spinors, $L_{\ell}^{i}$ the left-handed lepton doublet, and $\tilde{H} \equiv i \sigma^2 H^*$. The Yukawa couplings are assumed to be flavor-universal for simplicity, and the pure gauge-singlet sector is postulated to be $CP$-symmetric; hence, the flavor-universal right-handed neutrino Yukawa coupling, $y_{N}$, is real.\footnote{Note that the $CP$-symmetry forbids an operator of the form $y_{N}' \pbrac{S - S^*} \bar{\mathcal{N}}^{i} \gamma_{5} \mathcal{N}^{i}$ (leading to a decay of the $\chi$~pseudoscalar) given the Majorana nature of $\mathcal{N}^{i}$, the mass matrix of which is assumed to be real and diagonal in the weak basis.} The complex Dirac neutrino Yukawa matrix, $Y^\nu_{ij}$, may be ignored for the rest of the discussion, due to its extremely small entities (of the order of the electron Yukawa coupling) \cite{Farzinnia:2013pga}. The flavor-universal Yukawa coupling, subsequently, gives rise to a degenerate mass term for the right-handed Majorana neutrino flavors
\begin{equation}\label{MN}
M_{N} = \sqrt2 \, y_{N} \, v_{\eta} = \sqrt2 \, y_{N} \, v_{\phi} \cot\omega \ .
\end{equation}

In summary, the current framework introduces five new free parameters \cite{Farzinnia:2013pga}, which, without loss of generality, may be expressed as
\begin{equation}\label{inputs}
\cbrac{\omega, M_\chi, M_N, \lambda_\chi, \lambda_m^-}  \ ,
\end{equation}
and the remaining parameters are given in terms of this set according to
\begin{equation}\label{paraIV2}
\begin{split}
\lambda_\phi&=3\frac{M_h^2}{v_{\phi }^2} \cos ^2\omega\ ,\quad
\lambda_{m}^+=-\frac{M_h^2}{v_{\phi }^2} \sin ^2\omega\ ,\quad
\lambda_{\eta}=3\frac{M_h^2}{v_{\phi }^2} \sin^2\omega\tan^2\omega\ , \\
\lambda_{\eta\chi}&=\left(2\frac{M_\chi^2}{v_{\phi }^2}-\lambda _{m}^-\right)\tan^2\omega\ ,\quad
y_N=\frac{M_N}{\sqrt{2}\, v_\phi}\tan\omega \ .
\end{split}
\end{equation}
As noted in \cite{Farzinnia:2014xia}, one observes from the above relations that the sign of the mixing angle, $\omega$, is immaterial; whence, one may consider $0\leq\sin \omega \leq 1$, without loss of generality. In contrast, the sign of the $\lambda^{-}_{m}$ parameter remains formally and phenomenologically relevant.

Finally one deduces, from \eqref{flatdir}, that $\lambda_{m}^{+}<0$ along the flat direction, allowing for a successful spontaneous breaking of the electroweak symmetry breaking, as anticipated by \eqref{muSM}. Moreover, the dynamically-generated mass term for the Higgs field \eqref{muSM} can be expressed in terms of the free parameters of the theory using \eqref{paraIV2}, which yields
\begin{equation}\label{muSMfin}
\mu_H^{2} = \frac{\lambda_{m}^{+}}{2}\, v_{\eta}^{2} = -\frac{1}{2} M_{h}^{2} \cos^{2}\omega \ , 
\end{equation}
with $M_{h} = 125$~GeV. Hence, the spontaneous electroweak symmetry breaking is realized for $\sin \omega < 1$, corresponding to a nonzero singlet VEV, $v_{\eta} \neq 0$; in particular, a very large singlet VEV ($\sin \omega \sim 0$, c.f. \eqref{hs}), produces a SM-like mass term for the Higgs doublet field. This is a direct consequence of the restriction imposed by the scale symmetry upon the mixing parameter $\lambda_{m}^{+}$ (c.f. \eqref{paraIV2})
\begin{equation}\label{lmp}
\lambda_{m}^{+} = -\frac{M_h^2}{v_{\phi }^2 + v_{\eta}^2} \ ,
\end{equation}
implying that the coupling between the electroweak and the singlet scales is proportional to the ratio of the two, also manifesting in the definition of the mixing angle \eqref{hs}.\footnote{Note that the relation \eqref{lmp} guarantees the stability of the electroweak scale against quadratic contributions from the singlet scale. In a general singlet-extended model, without a protective symmetry, no such relation exists; the singlet VEV as well as the magnitude of its mixing with the electroweak sector are both independent free parameters, and require fine-tuning in order to prevent a quadratic destabilization of the electroweak scale.}

\subsection{One-loop Effective Scalar Potential}

At this point, let us examine the one-loop effective potential along the flat direction. Following \cite{Gildener:1976ih}, one may express this potential as
\begin{equation}\label{V1}
V^{(1)} (\varphi) = A\, \varphi^{4} + B\, \varphi^{4} \log \frac{\varphi^{2}}{\Lambda^{2}} \ ,
\end{equation}
with $A$ and $B$ dimensionless constants, parametrizing the contributions of all massive states within the loop, and $\varphi$ the radial combination of the $CP$-even scalars
\begin{equation}\label{varphi}
\varphi^{2} \equiv \phi^{2} + \eta^{2} \ .
\end{equation}
Note that the one-loop potential is defined at a definite energy scale, $\Lambda$, as anticipated.\footnote{Rewriting the tree-level potential in terms of the $\varphi$~field \eqref{varphi} along the flat direction \eqref{flatdir}, and using the definition of the mixing angle \eqref{hs}, one can show that the tree-level potential vanishes along the flat direction, leaving the one-loop effective potential \eqref{V1} as the dominant contribution along this direction.} The scale $\Lambda$ is found by minimizing the potential \eqref{V1} with respect to its constituent field $\varphi$, resulting in
\begin{equation}\label{Lamb}
\frac{d V^{(1)}}{d \varphi} \Big |_{\varphi = v_{\varphi}} = 0 \qquad \Longrightarrow \qquad \Lambda = v_{\varphi}\, \exp\tbrac{\frac{A}{2B} + \frac{1}{4}} \ .
\end{equation}
Inserting the expression for $\Lambda$ back into \eqref{V1} yields, subsequently, the simple expression for the one-loop effective potential
\begin{equation}\label{V1simp}
V^{(1)} (\varphi) = B\, \varphi^{4} \pbrac{ \log \frac{\varphi^{2}}{v_{\varphi}^{2}} - \frac{1}{2}} \ .
\end{equation}
This potential induces a radiative mass for the $\sigma$~boson, which is computed according to
\begin{equation}\label{msvarphi}
m_{\sigma}^{2} = \frac{d^{2} V^{(1)}}{d \varphi^{2}} \Big |_{\varphi = v_{\varphi}} = 8B \, v_{\varphi}^{2} \ .
\end{equation}

To find the one-loop effective potential of the SM $\phi$~field, one may project \eqref{V1simp} along its $\phi$~direction using \eqref{varphi} and \eqref{hs}, $\varphi^{2} = \phi^{2} / \sin^{2}\omega$,
\begin{equation}\label{V1phi}
V^{(1)} (\phi) = \beta\, \phi^{4} \pbrac{ \log \frac{\phi^{2}}{v_{\phi}^{2}} - \frac{1}{2}} \ ,
\end{equation}
and the minimization scale, $\Lambda$, takes the form \cite{Farzinnia:2013pga}
\begin{equation}\label{Lambfin}
\Lambda = v_{\phi}\, \exp\tbrac{\frac{\alpha}{2\beta} + \frac{1}{4}} \ .
\end{equation}
The coefficients $\alpha$ and $\beta$ are related to the original variables $A$ and $B$ \cite{Farzinnia:2013pga}, and are defined in the $\overline{\text{MS}}$-scheme by
\begin{align}
\alpha = &\, \frac{1}{64\pi^2 v_\phi^4} \Bigg[ M_h^4\pbrac{-\frac{3}{2}+\log\frac{M_h^2}{v_\phi^2}}
  + M_\chi^4\pbrac{-\frac{3}{2}+\log\frac{M_\chi^2}{v_\phi^2}} + 6M_W^4\pbrac{-\frac{5}{6}+\log\frac{M_W^2}{v_\phi^2}} \notag \\
& \qquad
   + 3M_Z^4\pbrac{-\frac{5}{6}+\log\frac{M_Z^2}{v_\phi^2}}
   - 12M_t^4\pbrac{-1+\log\frac{M_t^2}{v_\phi^2}}
   - 6M_N^4\pbrac{-1+\log\frac{M_N^2}{v_\phi^2}} \Bigg] \ , \label{alpha} \\
\beta =&\, \frac{1}{64\pi^2 v_\phi^4} \pbrac{M_h^4 + M_\chi^4 + 6M_W^4+3M_Z^4 -12 M_t^4 - 6 M_N^4} \label{beta}\ .
\end{align}
In this expression, the contribution of all the relevant massive particles in the loop, including the heavy SM states, is taken into account, with the numerical coefficients representing the internal degree of freedom associated with each particle species. The radiative mass of the $\sigma$~boson is then given by \cite{Farzinnia:2013pga}
\begin{equation}\label{ms}
m_{\sigma}^{2} = 8\beta \, v_{\phi}^{2} \sin^{2}\omega \ .
\end{equation}
As a consequence, in our analysis, one may trade either of the first three input parameters in \eqref{inputs} by the mass of the $\sigma$~boson, as an alternative free parameter of the model.

One notes that the one-loop effective potential \eqref{V1phi} is bounded from below for $\beta >0$, which, at the same time, guarantees that the $\sigma$~boson does not become tachyonic (c.f. \eqref{ms}). This requirement results in the nontrivial relation between the masses 
\begin{equation}\label{staboneloop}
M_\chi^4 - 6 M_N^4 > 12 M_t^4 - 6M_W^4 - 3M_Z^4 - M_h^4 \ .
\end{equation}
It is evident that this relation cannot be satisfied within the SM alone, indicating the failure of the Coleman-Weinberg mechanism within the pure SM.

\section{Discovery Prospects from the LHC and the XENON1T Experiments}\label{prosp}

Having reviewed the formal anatomy of the model in Sec.~\ref{rev}, we proceed to study the strongest projected exclusion limits on the parameter space from the ATLAS at $\sqrt s =14$~TeV in the vector boson channels with an integrated luminosity of 300 and 3000~fb$^{-1}$ \cite{ATLASHeavyS}, as well as the XENON1T dark matter direct detection experiment \cite{Aprile:2012zx}. As discussed, the model contains two Higgs-like scalars, both of which are capable of interacting with the SM content; namely, the $h$ and $\sigma$~bosons, where the $h$~scalar has been identified with the discovered boson at the LHC \cite{LHCnew}, $M_{h} = 125$~GeV. The LHC prospects for discovering a heavy Higgs-like boson may then be applied to the $\sigma$~scalar, in order to study the projected constraints on its properties.\footnote{In principle, the $\sigma$~boson may be heavier or lighter than the $h$~scalar. In light of the considered ATLAS heavy Higgs projections, in this treatment, we focus specifically on the case where the $\sigma$~boson is heavier than the $h$~scalar.} Furthermore, the pseudoscalar $\chi$ of the scenario is a stable WIMP dark matter candidate, due to the $CP$-symmetry of the potential, and the discovery prospects from the XENON1T direct detection experiment may be additionally utilized to analyze the parameter space.

To investigate the properties of the $\sigma$~boson using the available 95\%~C.L. LEP \cite{LEPII} and the $\sqrt s =7,8$~TeV LHC \cite{LHCHeavyH} data, an effective Lagrangian was constructed in \cite{Farzinnia:2014xia}, parametrizing the deviations of the $\sigma$~boson couplings to the (kinematically) available decay pairs from their corresponding SM values\footnote{As discussed in \cite{Farzinnia:2014xia}, a decay of the $\sigma$~boson into a pair of the $\chi$~pseudoscalars is kinematically forbidden for $m_{\sigma}\leq 1$~TeV.}
\begin{equation}\label{Leff}
\begin{split}
\mathcal{L}^\sigma_{\text{effective}} =&\,
   c^\sigma_V \frac{2M_W^2}{v_\phi}\,\sigma\,W_{\mu}^{+}W^{-\mu}
  +c^\sigma_V \frac{M_Z^2}{v_\phi}\,\sigma\,Z_{\mu}Z^{\mu}
  -c^\sigma_t \frac{M_t}{v_\phi}\,\sigma\,\bar{t}t
  -c^\sigma_b \frac{M_b}{v_\phi}\,\sigma\,\bar{b}b
  -c^\sigma_c \frac{M_c}{v_\phi}\,\sigma\,\bar{c}c
  -c^\sigma_\tau \frac{M_\tau}{v_\phi}\,\sigma\,\bar{\tau}\tau\\
&
  +c^\sigma_g\frac{\alpha_s}{12\pi{v_\phi}}\,\sigma\,G^a_{\mu\nu}G^{a\, \mu\nu}
  +c^\sigma_\gamma\frac{\alpha}{\pi{v_\phi}}\,\sigma\,A_{\mu\nu}A^{\mu\nu}
  +c^\sigma_h \,\sigma\,h h
  +c^\sigma_\mathcal{N} \,\sigma\,\bar{\mathcal{N}}^{i}\mathcal{N}^{i} \ ,
  \end{split}
\end{equation}
where ($\phi$ representing the SM Higgs boson), the coupling coefficients assume the following values within the current framework
\begin{equation}\label{Leffcoeff}
\begin{split}
&c^\sigma_V=c^\sigma_t=c^\sigma_b=c^\sigma_c=c^\sigma_\tau=\sin\omega \ , \qquad c^\sigma_g= \sin \omega\times c_g^{\phi}(m_{\phi} = m_{\sigma}) \ , \qquad c^\sigma_\gamma= \sin \omega\times c_\gamma^{\phi}(m_{\phi} = m_{\sigma}) \ , \\
&c^\sigma_h = - \frac{M_{h}^{2}}{v_{\phi}}\sin \omega \ , \qquad c^\sigma_\mathcal{N} = - \frac{M_{N}}{2v_{\phi}}\sin \omega \ .
  \end{split}
\end{equation}

The same effective Lagrangian \eqref{Leff} is suitable for analyzing the ATLAS heavy SM-like Higgs projections in the vector boson channels \cite{ATLASHeavyS}, applied to the $\sigma$~boson. In analogy with the heavy Higgs data analysis at $\sqrt s =7,8$~TeV, for the current study, we construct the ``$\mu$''~parameter. The latter, using the narrow-width approximation,\footnote{It has been shown in \cite{Farzinnia:2014xia} that the narrow-width approximation remains valid for the $\sigma$~scalar up to $m_{\sigma}\leq 1$~TeV.} is defined as the product of the $\sigma$~boson production cross section and its vector boson decay branching ratios, divided by those of a purely SM-like scalar with the same mass
\begin{equation}\label{mupar}
\mu(ii \to \sigma \to VV)\equiv \frac{\sigma(ii \to \sigma) \times \textrm{BR}(\sigma \to VV)}{\sigma(ii \to \phi) \times \textrm{BR}(\phi \to VV)} = \sin^4\omega \, \frac{\Gamma^{\phi}_\text{total}(m_{\phi}=m_\sigma)}{\Gamma_\text{total}^{\sigma}} \ ,
\end{equation}
where, $ii$ denotes the vector boson fusion, vector-Higgs associated production, and the gluon-fusion channels. In the final quantity on the right-hand side of \eqref{mupar}, the total decay width of the $\sigma$~scalar is computed according to \cite{Farzinnia:2014xia}
\begin{equation}\label{Ctots}
\begin{split}
\Gamma_\text{total}^{\sigma} =&\, \sin^2\omega\tbrac{\textrm{BR}^{\textrm{SM}}_{WW} + \textrm{BR}^{\textrm{SM}}_{ZZ} + \textrm{BR}^{\textrm{SM}}_{gg} + \textrm{BR}^{\textrm{SM}}_{\gamma \gamma}+\textrm{BR}^{\textrm{SM}}_{\bar{t}t}+\textrm{BR}^{\textrm{SM}}_{\bar{b}b}+\textrm{BR}^{\textrm{SM}}_{\bar{c}c}+\textrm{BR}^{\textrm{SM}}_{\bar{\tau}\tau}} \Gamma^{\phi}_\text{total}(m_{\phi}=m_\sigma) \\
&+\sin^{2}\omega\,\frac{M_{h}^4}{8\pi v_{\phi}^{2} \,m_\sigma}\sqrt{1-\pbrac{\frac{2M_{h}}{m_\sigma}}^{2}}+ \sin^{2}\omega\,\frac{m_\sigma M_{N}^2}{16\pi v_{\phi}^{2}}\tbrac{1-\pbrac{\frac{2M_{N}}{m_\sigma}}^{2}}^{3/2} \ ,
\end{split}
\end{equation}
with the two expressions on the last line of \eqref{Ctots} representing the decay widths of the $\sigma$~boson into a pair of the $h$~scalars and a pair of the right-handed Majorana neutrinos, respectively.

Although the $\sqrt s =7,8$~TeV heavy Higgs searches \cite{LHCHeavyH} provide the $\mu$~parameter directly for the vector boson decay channels, the ATLAS $\sqrt s =14$~TeV heavy Higgs projections \cite{ATLASHeavyS} quote separately the $\sigma$~scalar production cross section times vector boson decay branching ratios, and those for the SM-like $\phi$~Higgs with the same mass (the most stringent projected bounds are due to the $ZZ$~decay channel). Hence, we compute the $\mu$~parameter for the ATLAS projections simply by dividing these two separately quoted quantities (c.f. the definition in \eqref{mupar}).

\begin{figure}
\includegraphics[width=.497\textwidth]{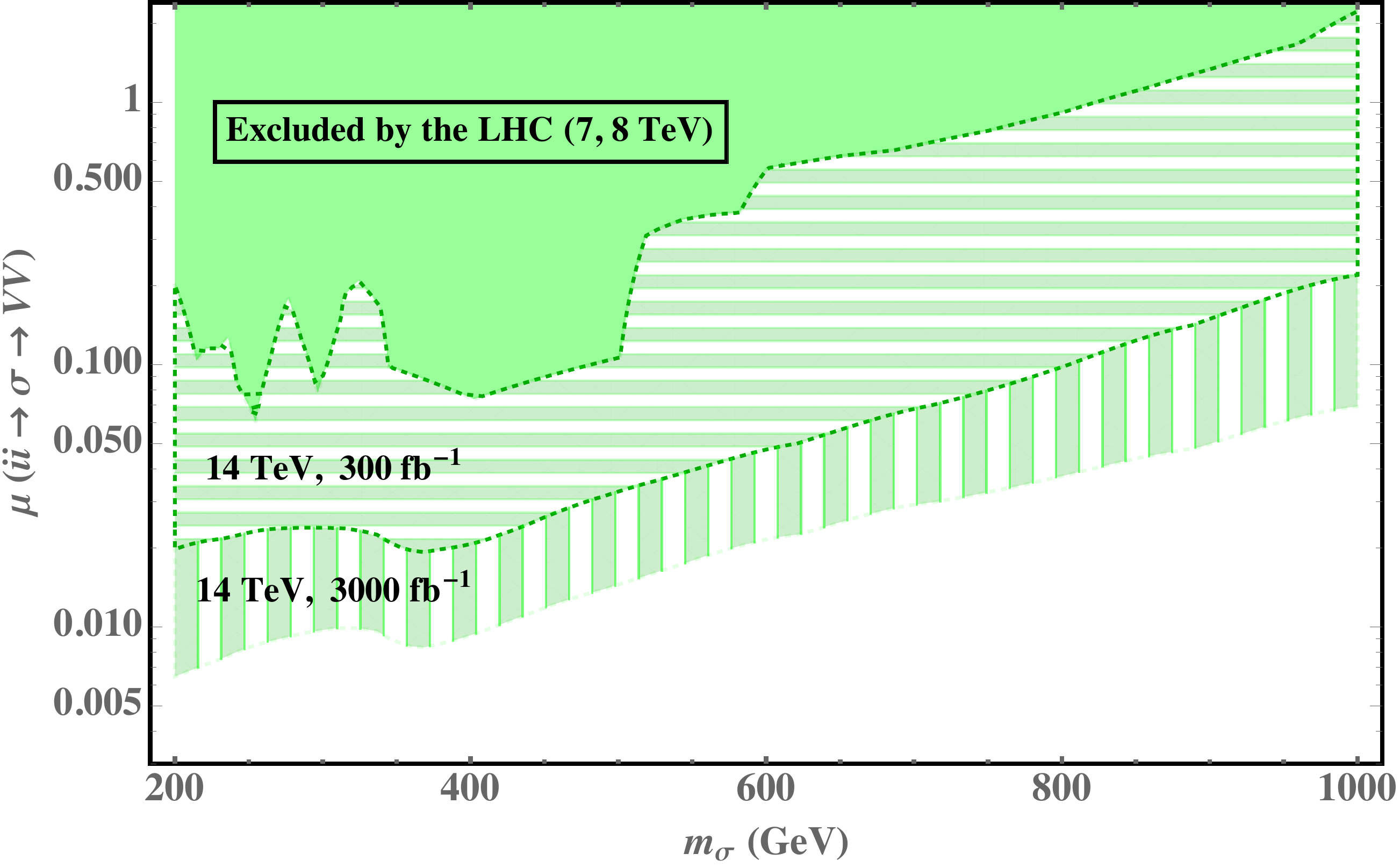}
\includegraphics[width=.497\textwidth]{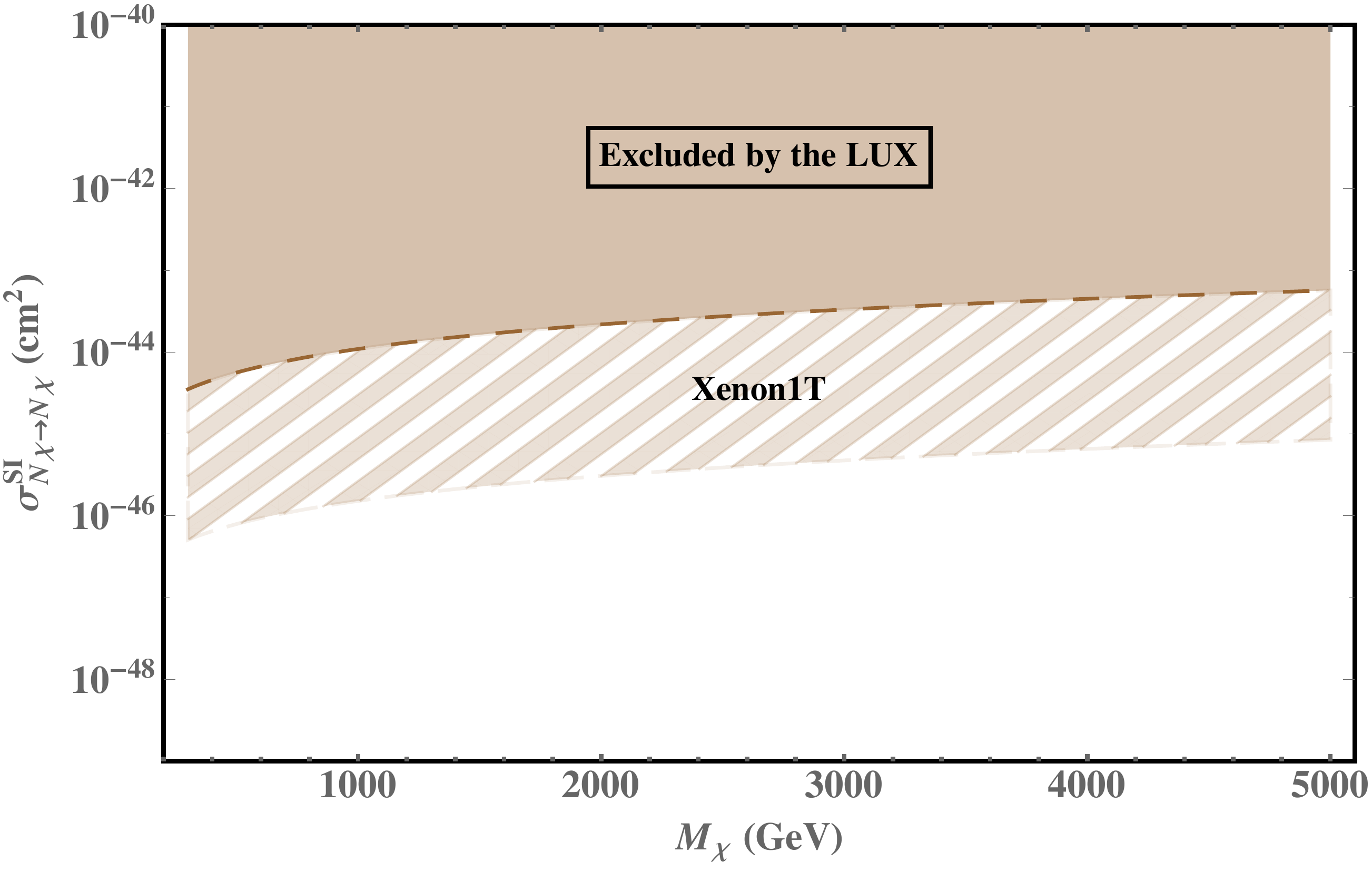}
\caption{\textit{Left}: The $\mu$~parameter \eqref{mupar}, defined as the product of the $\sigma$~boson production cross section and its vector boson decay branching ratios divided by those of a purely SM-like boson with the same mass, as a function of the $\sigma$~boson mass for the range $200 \leq m_{\sigma} \leq 1000$~GeV. The solid green region is excluded by the most stringent LHC $\sqrt s = 7,8$~TeV Higgs searches at 95\%~C.L.~\cite{LHCHeavyH}. In case of no LHC heavy Higgs discovery at $\sqrt s = 14$~TeV with an integrated luminosity of 300~fb$^{-1}$ (3000~fb$^{-1}$) the horizontally (vertically) striped green region is expected to be additionally excluded by the ATLAS projections at 95\%~C.L.~\cite{ATLASHeavyS}. \textit{Right}: The spin-independent WIMP-nucleon scattering cross section as a function of the WIMP mass for the range $100 \lesssim M_{\chi} \leq 5000$~GeV. The solid brown region is excluded by the LUX direct detection experiment at 90\%~C.L.~\cite{LUX2013}. In case of no dark matter signal discovery by the XENON1T experiment, the inclined striped brown region is expected to be additionally excluded at 90\%~C.L.~\cite{Aprile:2012zx}. (The same displayed color-coding in both panels will be used throughout this treatment)}
\label{LHCDM}
\end{figure}

The left panel of Fig.~\ref{LHCDM} displays, in the $m_{\sigma}-\mu$~plane for $200 \leq m_{\sigma} \leq 1000$~GeV, the most stringent exclusion bounds by the LHC $\sqrt s = 7,8$~TeV Higgs searches at 95\%~C.L.~\cite{LHCHeavyH}, as well as the ATLAS projections for the complementary constraints at $\sqrt s = 14$~TeV with an integrated luminosity of 300~fb$^{-1}$  and 3000~fb$^{-1}$ at 95\%~C.L.~\cite{ATLASHeavyS}, in case no heavy Higgs-like boson is discovered. We shall compare the theoretically constructed $\mu$~parameter of the model \eqref{mupar} with the displayed ATLAS heavy Higgs discovery prospects and analyze the projected constraints on the parameter space.

Next, we consider the expected reach of the XENON1T direct detection experiment \cite{Aprile:2012zx} for the spin-independent scattering cross section of the dark matter particles off the nucleons. An analysis of the direct detection exclusion limits of the LUX experimental data at 90\%~C.L. \cite{LUX2013}, applied to the $\chi$~pseudoscalar, was previously performed in \cite{Farzinnia:2014xia}. As explained in \cite{Farzinnia:2014xia}, the heavy $\chi$~WIMP of the scenario interacts with the nucleons via the $t$-channel exchange of the $h$~and~$\sigma$~bosons, and the spin-independent interaction cross section was calculated as
\begin{equation}\label{DMdirCS}
\sigma^{\text{SI}}_{N\chi \to N\chi} =\frac{g_{W}^{2}}{16\pi} \, \frac{m_{N}^{4}f_{N}^{2}}{M_{W}^{2} M_{\chi}^{2}} \tbrac{\frac{\lambda_{\chi\chi h}}{M_{h}^{2}}\cos \omega + \frac{\lambda_{\chi\chi\sigma}}{m_{\sigma}^{2}}\sin \omega}^{2} \ ,
\end{equation}
with $g_{W}$ the weak coupling, $m_{N} = 0.939$~GeV the average nucleon mass, and $f_{N} \simeq 0.345$ \cite{Cline:2013gha,Agrawal:2010fh,Ellis:2000ds} the nucleon form factor.\footnote{See also \cite{Crivellin:2013ipa} for further discussions addressing the caveats related to the nucleon form factor.} The coefficients $i\lambda_{\chi\chi h}$ and $i\lambda_{\chi\chi\sigma}$ represent, respectively, the couplings of the $h$~and~$\sigma$~scalars to the dark matter pair and were provided in \cite{Farzinnia:2014xia}. A destructive interference between the two scalar $t$-channel interactions may occur in \eqref{DMdirCS} for suitable values of the parameters, alleviating the direct detection bounds.

The right panel of Fig.~\ref{LHCDM} depicts, for $100 \lesssim M_{\chi} \leq 5000$~GeV, the spin-independent WIMP-nucleon scattering cross section as a function of the WIMP mass, incorporating the experimental upper bound from the available LUX data at 90\%~C.L.~\cite{LUX2013}, as well as the expected complementary limits from the XENON1T experiment at 90\%~C.L.~\cite{Aprile:2012zx}, in case of a null dark matter direct detection result.\footnote{For illustrative purposes, the XENON1T prospects are extrapolated to larger dark matter values (see also \url{http://dendera.berkeley.edu/plotter/entryform.html}).} We shall extend the experimental limits on the parameter space previously obtained using the LUX direct detection data, by making a comparison between the model's derived spin-independent scattering cross section \eqref{DMdirCS} and the expected discovery reach of the XENON1T direct detection experiment, and analyze the consequences for the free parameter space. 

At this stage, let us summarize the discussed collider and dark matter projection constraints within the exclusion plots for various benchmark values of the free parameters \eqref{inputs}. Moreover, we incorporate in the same plots the limits arising from considering perturbative unitarity, electroweak precision tests, and the LHC direct measurements of the 125~GeV Higgs properties, previously studied in \cite{Farzinnia:2013pga}, as well as the LHC $\sqrt s =7,8$~TeV heavy Higgs searches \cite{LHCHeavyH}, the LUX direct detection exclusion limits \cite{LUX2013}, and the (updated) WIMP thermal relic abundance determined by the Planck collaboration \cite{Ade:2015xua}, previously analyzed in \cite{Farzinnia:2014xia}. Since none of the studied constraints depends on the dark matter quartic self-coupling, $\lambda_{\chi}$, the latter may be ignored as an input for the rest of this discussion, and one may focus on the remaining four free parameters, $\omega$, $M_{\chi}$, $M_{N}$, and $\lambda_{m}^{-}$. Note that, as mentioned in Sec.~\ref{rev}, any of the $\omega$, $M_{\chi}$, and $M_{N}$ parameters may be replaced by $m_{\sigma}$, according to its definition \eqref{ms}.

\begin{figure}
\includegraphics[width=.329\textwidth]{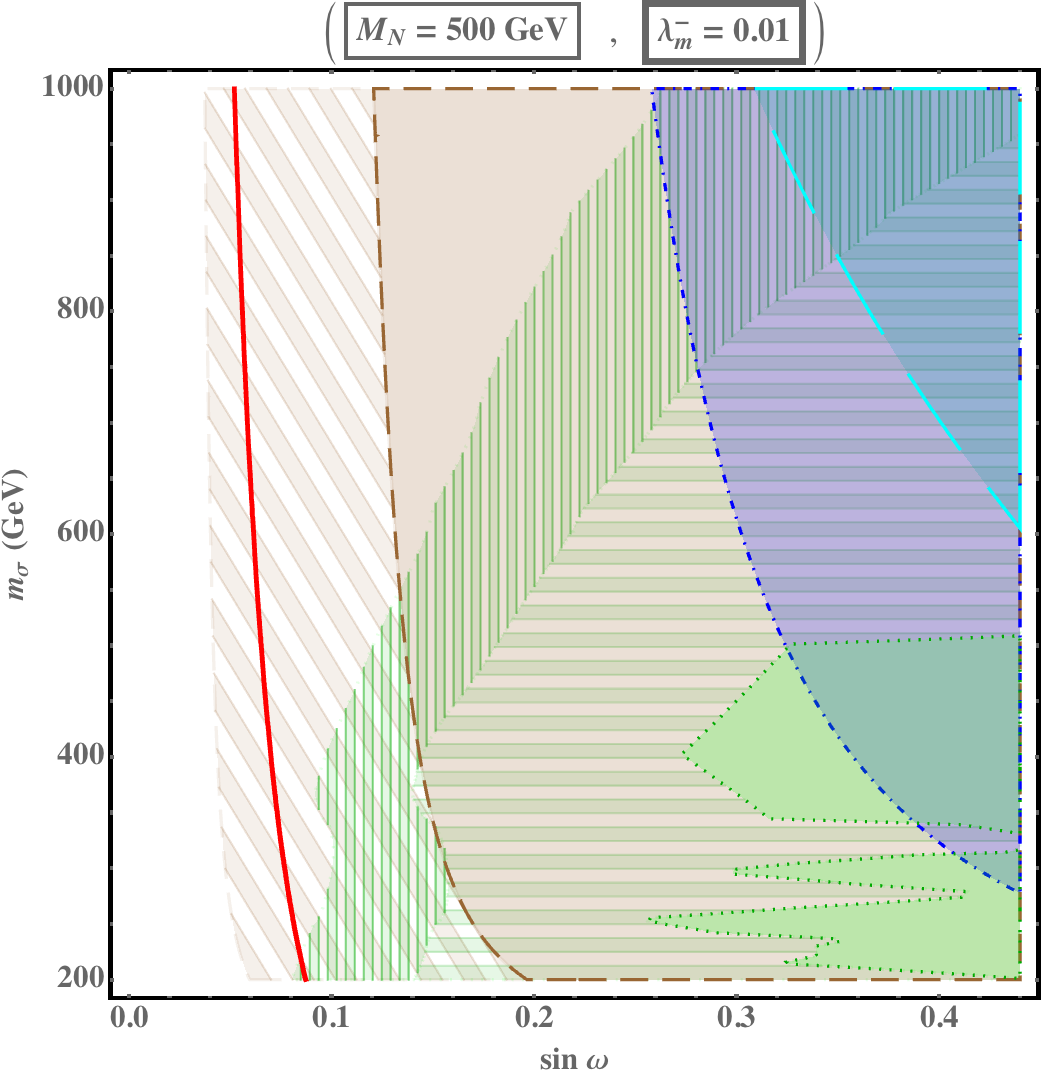}
\includegraphics[width=.329\textwidth]{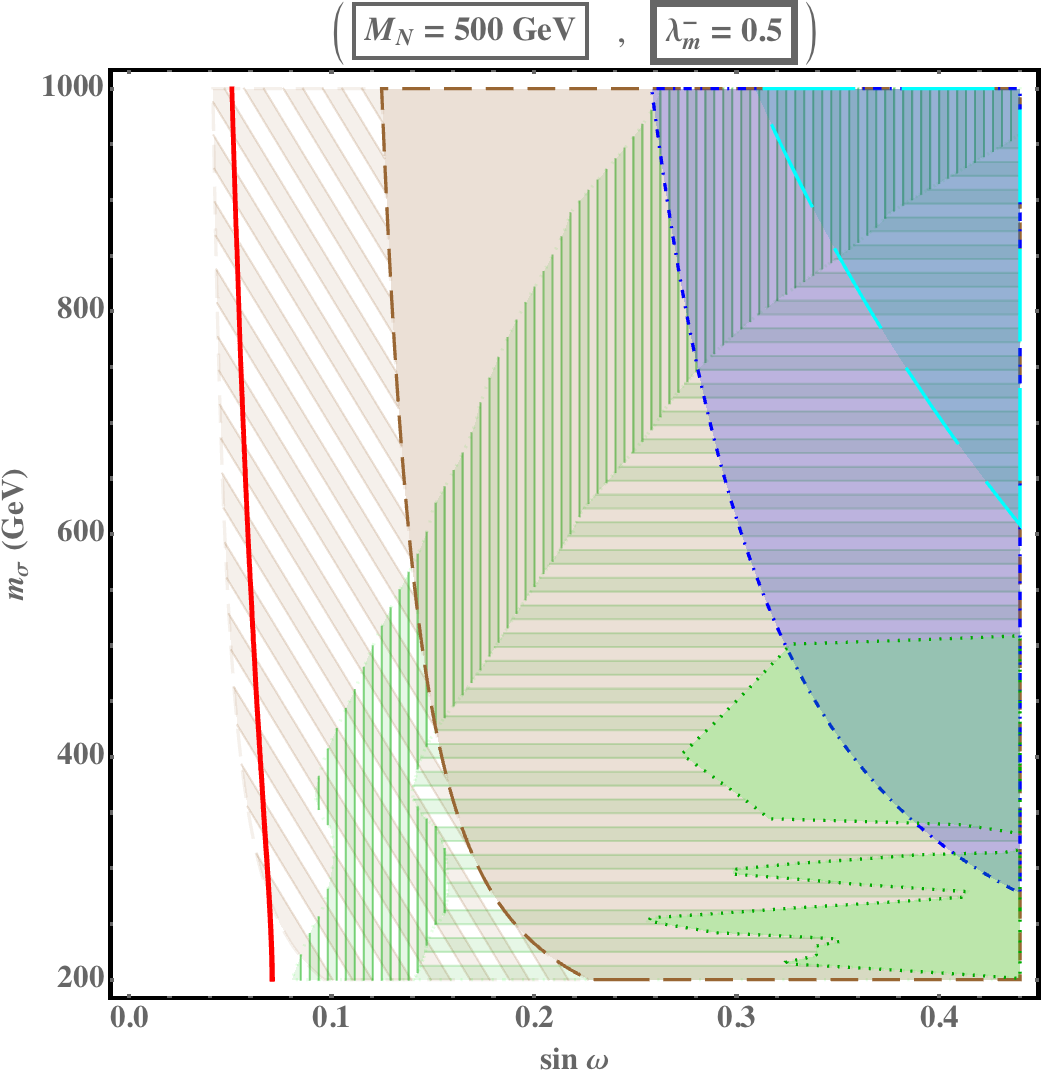}
\includegraphics[width=.329\textwidth]{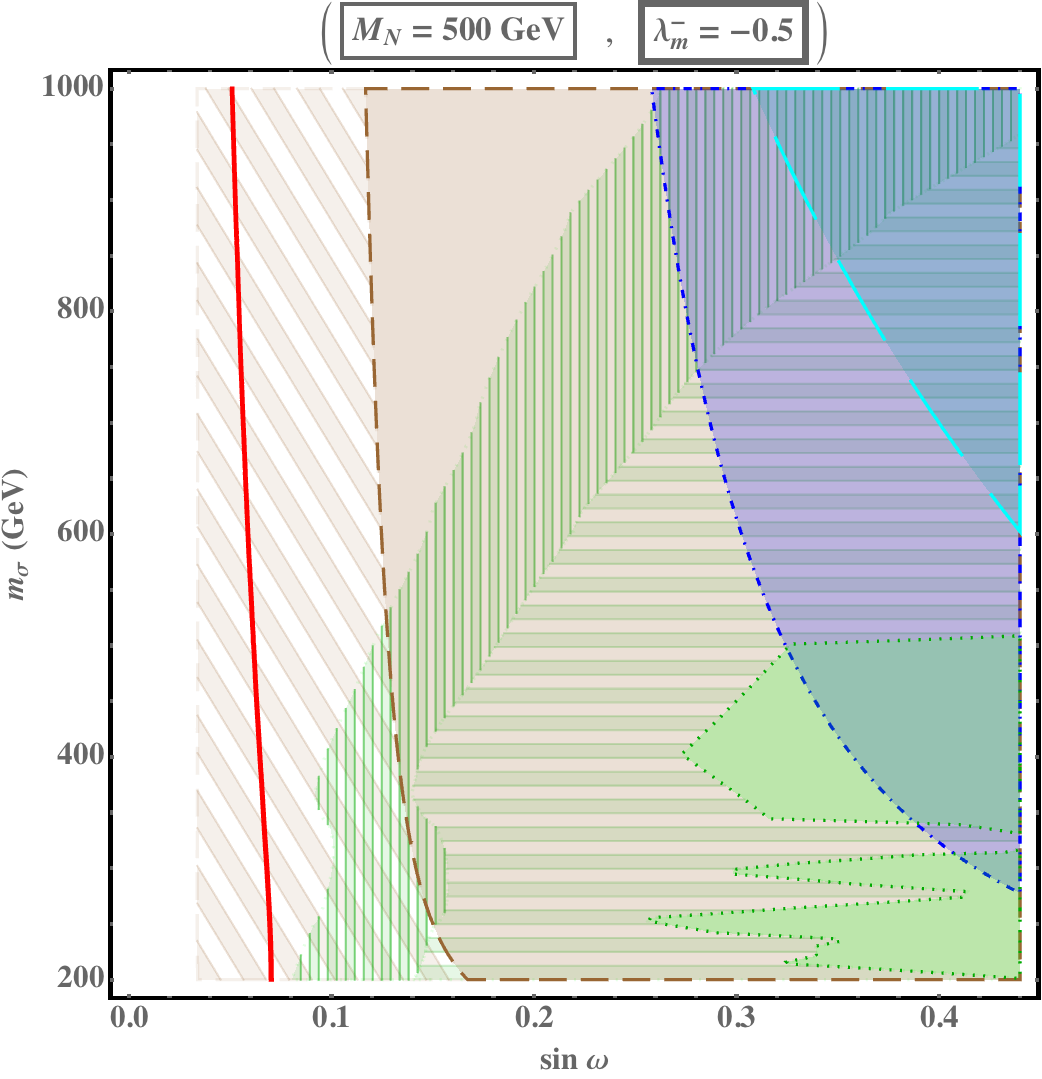}
\includegraphics[width=.329\textwidth]{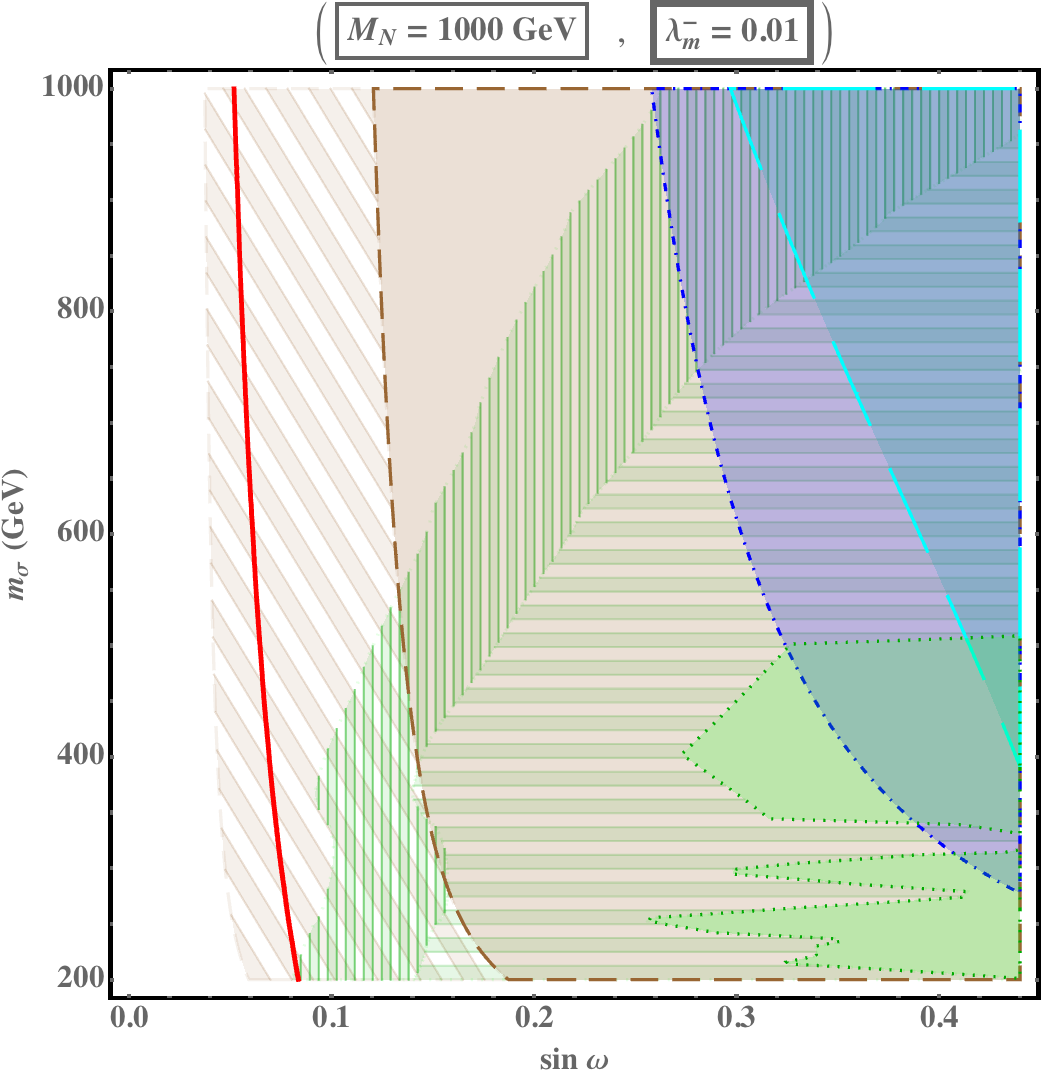}
\includegraphics[width=.329\textwidth]{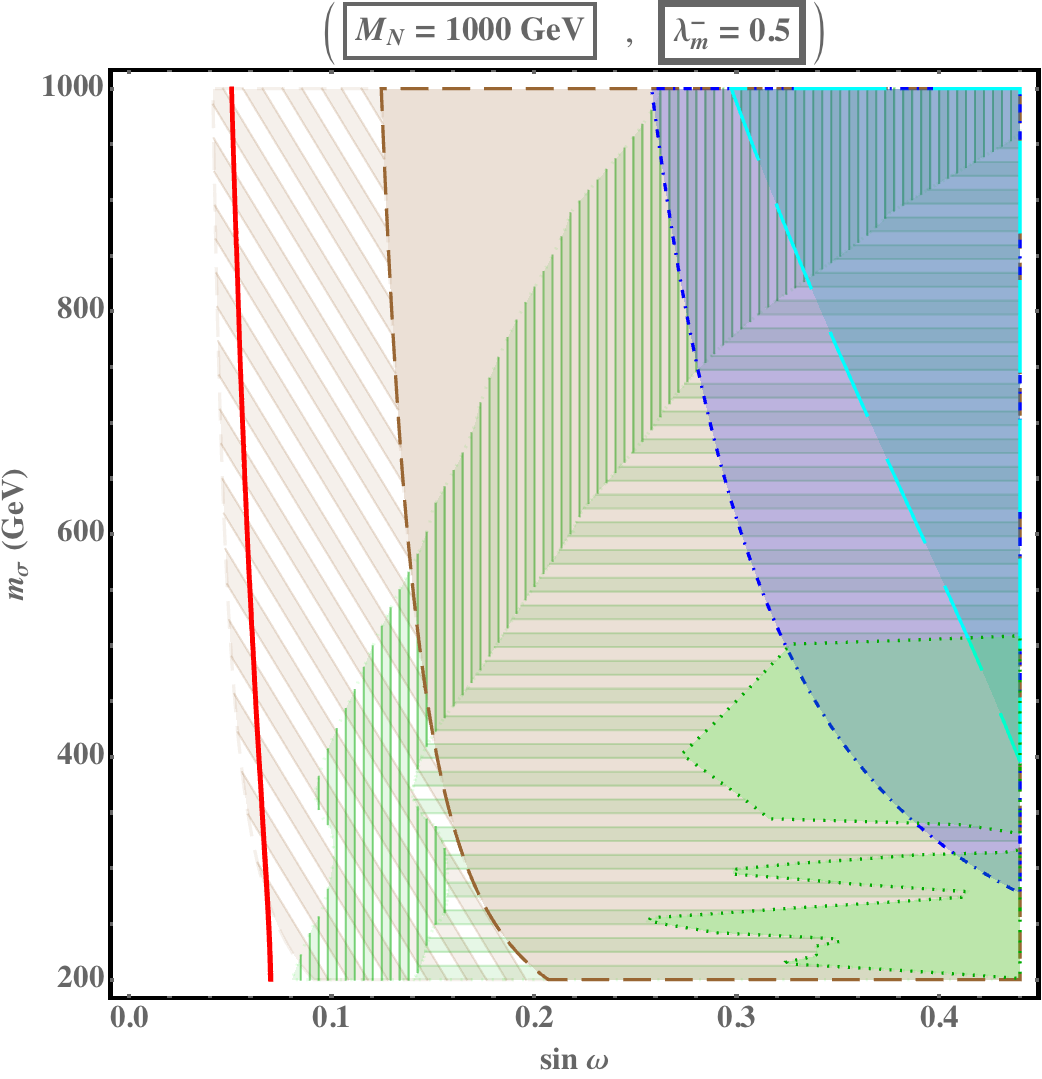}
\includegraphics[width=.329\textwidth]{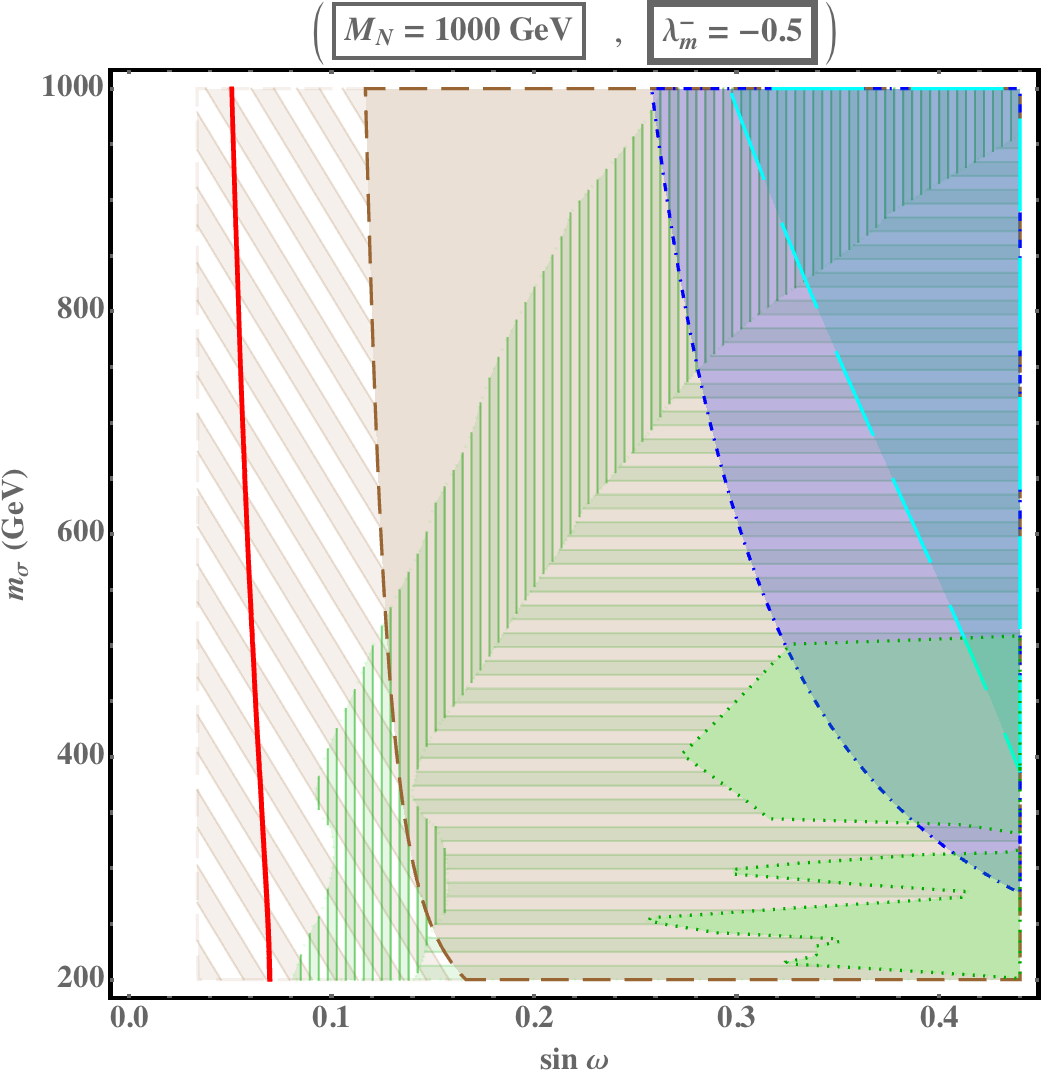}
\includegraphics[width=.329\textwidth]{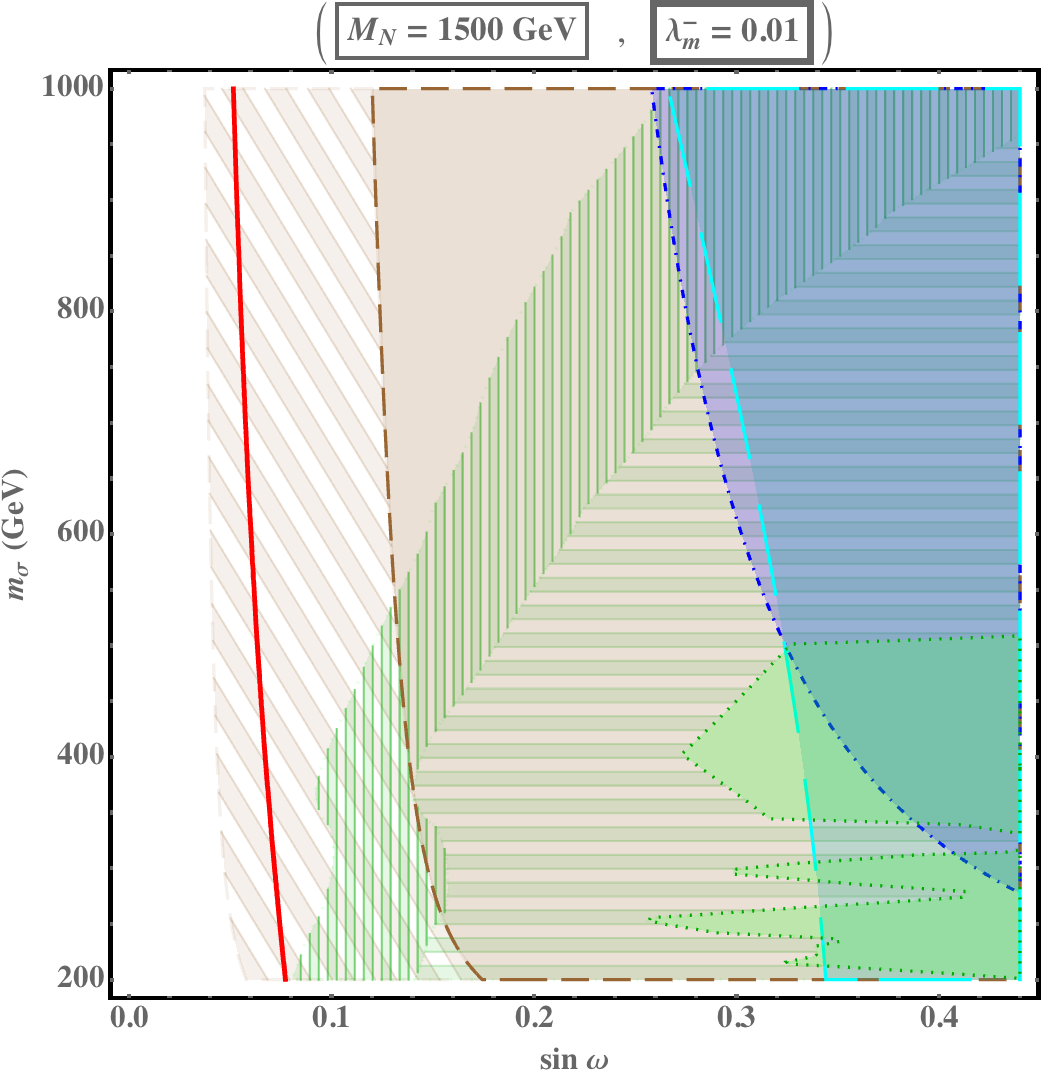}
\includegraphics[width=.329\textwidth]{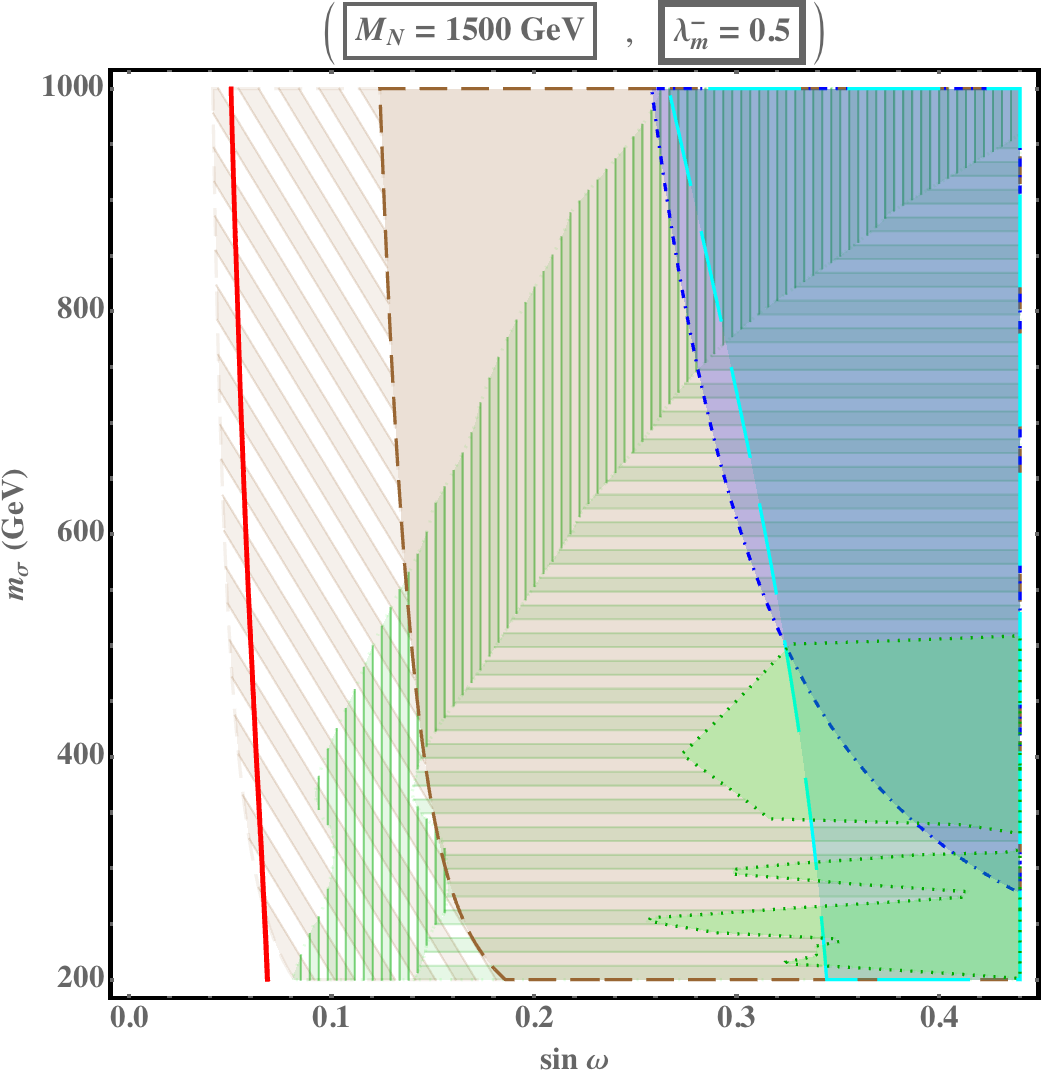}
\includegraphics[width=.329\textwidth]{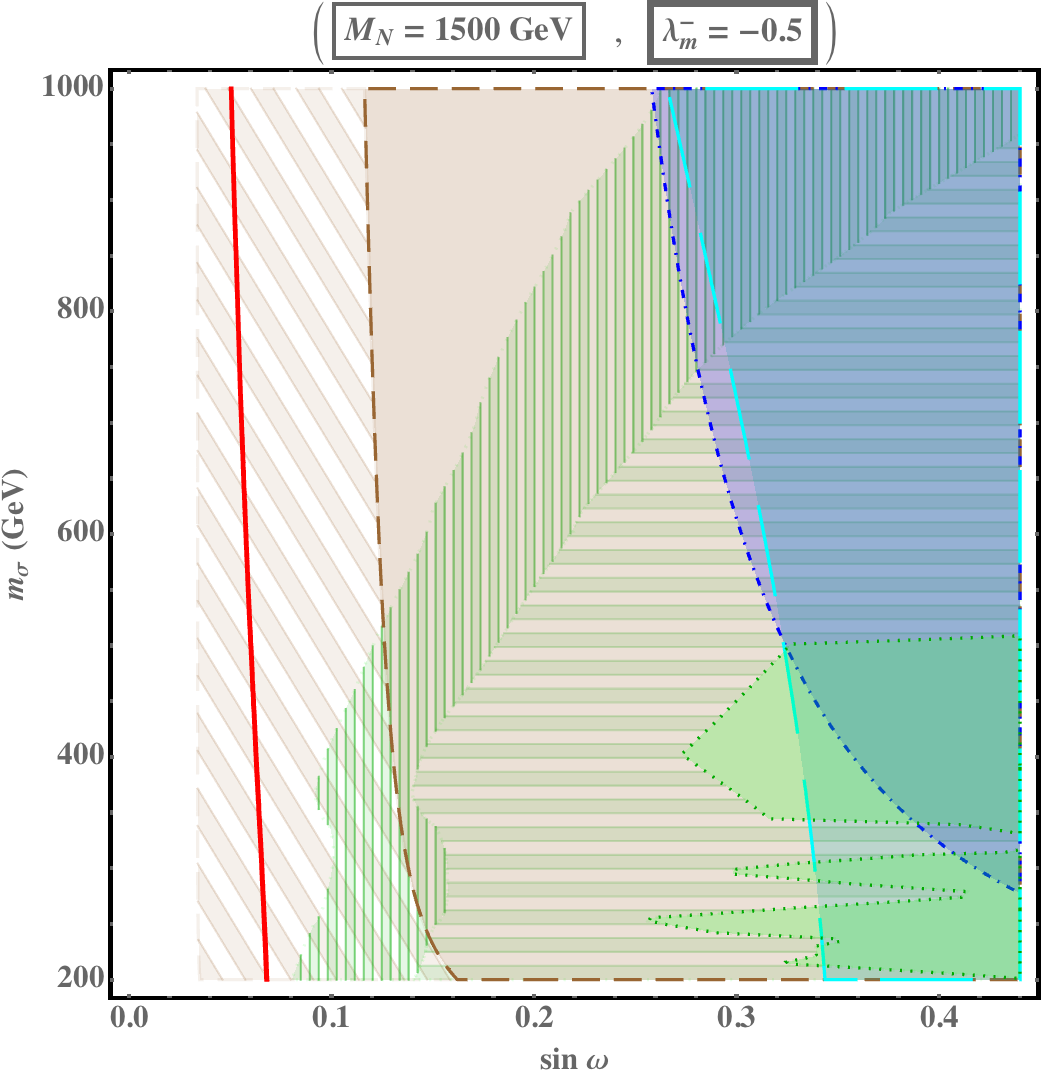}
\caption{Theoretical and experimental limits on the $\sin\omega-m_{\sigma}$~plane, for benchmark values of the remaining two input parameters, $M_{N}$ and $\lambda_{m}^{-}$. Note that the latter may assume a positive as well as a negative value. All solid-colored regions are excluded. The LHC direct measurements of the 125~GeV Higgs boson properties constrain the mixing angle to $\sin\omega \lesssim 0.44$ at 95\%~C.L., whereas the LHC heavy Higgs searches and projections are limited to the range $200 \leq m_{\sigma} \leq 1000$~GeV, which define the range of the plotted axes. The depicted exclusion bounds arise due to the perturbative unitarity (long-dashed blue line), the electroweak precision tests at 95\%~C.L. (dot-dashed purple line), the LHC heavy Higgs searches at 95\%~C.L. (dotted green line), and the LUX direct detection data at 90\%~C.L. (short-dashed brown line). The solid vertical red line represents the thermal relic abundance as reported by the Planck collaboration, and its thickness corresponds to the quoted $1\sigma$~uncertainty. Furthermore, in the event of a null signal discovery, the horizontally (vertically) striped green region becomes additionally excluded by the $\sqrt s =14$~TeV ATLAS heavy Higgs projections with an integrated luminosity of 300~fb$^{-1}$ (3000~fb$^{-1}$) at 95\%~C.L., while the inclined striped brown region becomes additionally excluded at 90\%~C.L. by the XENON1T prospects. The mass of the right-handed Majorana neutrino, $M_{N}$, affects most prominently the perturbative unitarity bound, whereas the magnitude as well as the sign of the $\lambda_{m}^{-}$~coupling have a dominant influence on the direct detection limits and projections.}
\label{sinwms}
\end{figure}

Fig.~\ref{sinwms} demonstrates the aforementioned obtained bounds within the $\sin\omega-m_{\sigma}$~exclusion plots for various choices of the remaining two input parameters, $M_{N}$ and $\lambda_{m}^{-}$. Since the $\lambda_{m}^{-}$~coupling may be positive or negative, benchmark values with both signs have been considered. The reported LHC heavy Higgs searches and projections cover the mass range $200 \leq m_{\sigma} \leq 1000$~GeV; therefore, the vertical axis is limited to these values. The LHC direct measurements of the 125~GeV Higgs boson properties, on the other hand, constrain the mixing angle to $\sin\omega \lesssim 0.44$ at 95\%~C.L. \cite{Farzinnia:2013pga,Farzinnia:2014xia}, which is reflected in the displayed domain of the horizontal axis.\footnote{As discussed in \cite{Farzinnia:2014xia}, the upper limit on the mixing angle $\sin\omega \lesssim 0.37$ reported in \cite{Farzinnia:2013pga} is incorrect, due to a minor error in the fitting code, and the correct limit corresponds to $\sin\omega \lesssim 0.44$ at 95\%~C.L.}

It is evident that the effect of the right-handed Majorana neutrino mass, $M_{N}$, is most prominently pronounced in the perturbative unitarity bound, with a heavier mass excluding a larger region of the parameter space. In contrast, the magnitude as well as the sign of the $\lambda_{m}^{-}$~quartic coupling largely affect the direct detection limits and projections, with a negative $\lambda_{m}^{-}$---corresponding to an attractive interaction between pairs of $\phi$ and pairs of $\chi$ (c.f. \eqref{V0quart})---resulting in a more stringent constraint. Taking into account the discussed ATLAS and XENON1T projections, in case no heavy Higgs-like scalar or pseudoscalar WIMP is discovered, the parameter space becomes increasingly more constrained, predominantly due to the expected XENON1T direct detection reach. The latter constrains the mixing angle to $\sin\omega \lesssim 0.04$, with a mild dependence on the mass of the $\sigma$~boson within the considered range and the values of the remaining input parameters.

\begin{figure}
\includegraphics[width=.329\textwidth]{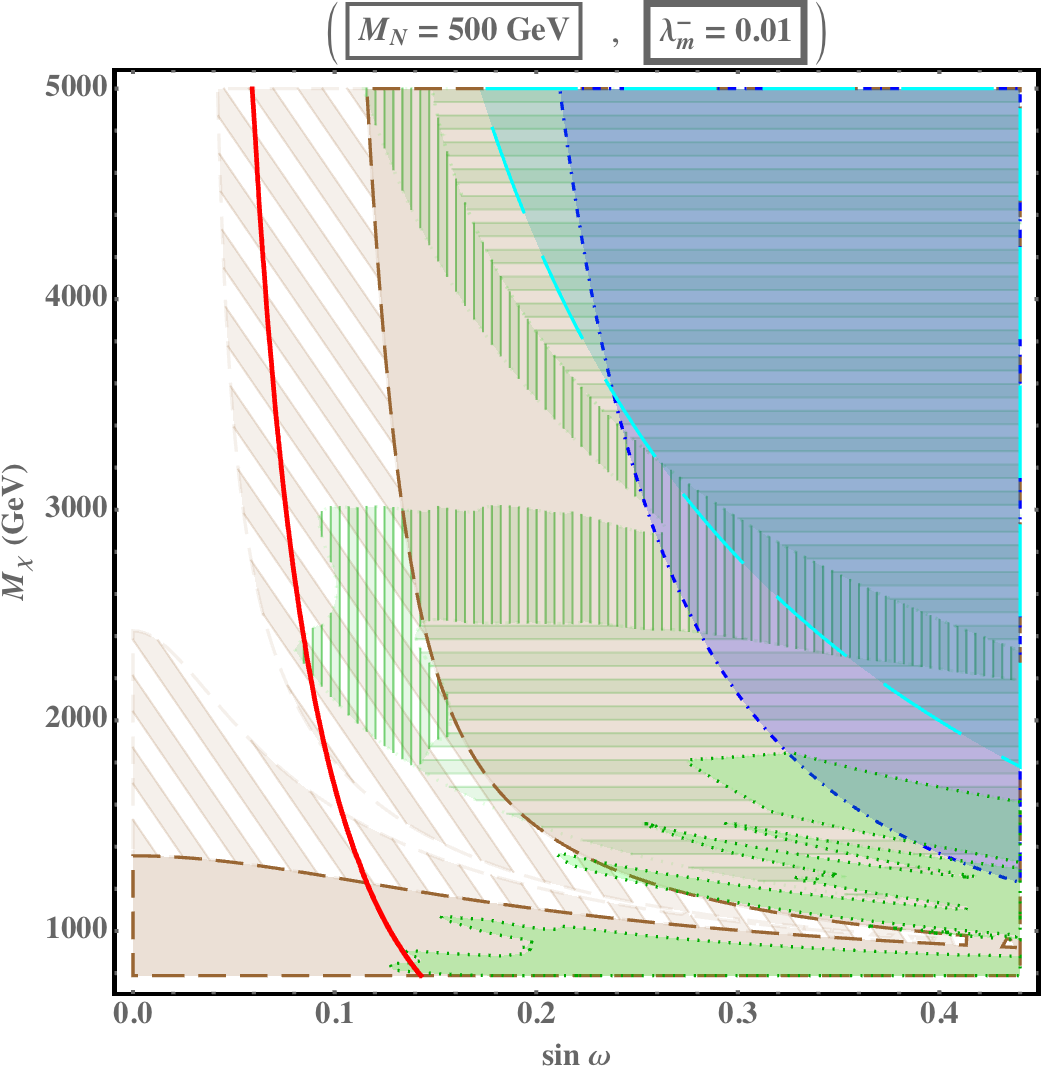}
\includegraphics[width=.329\textwidth]{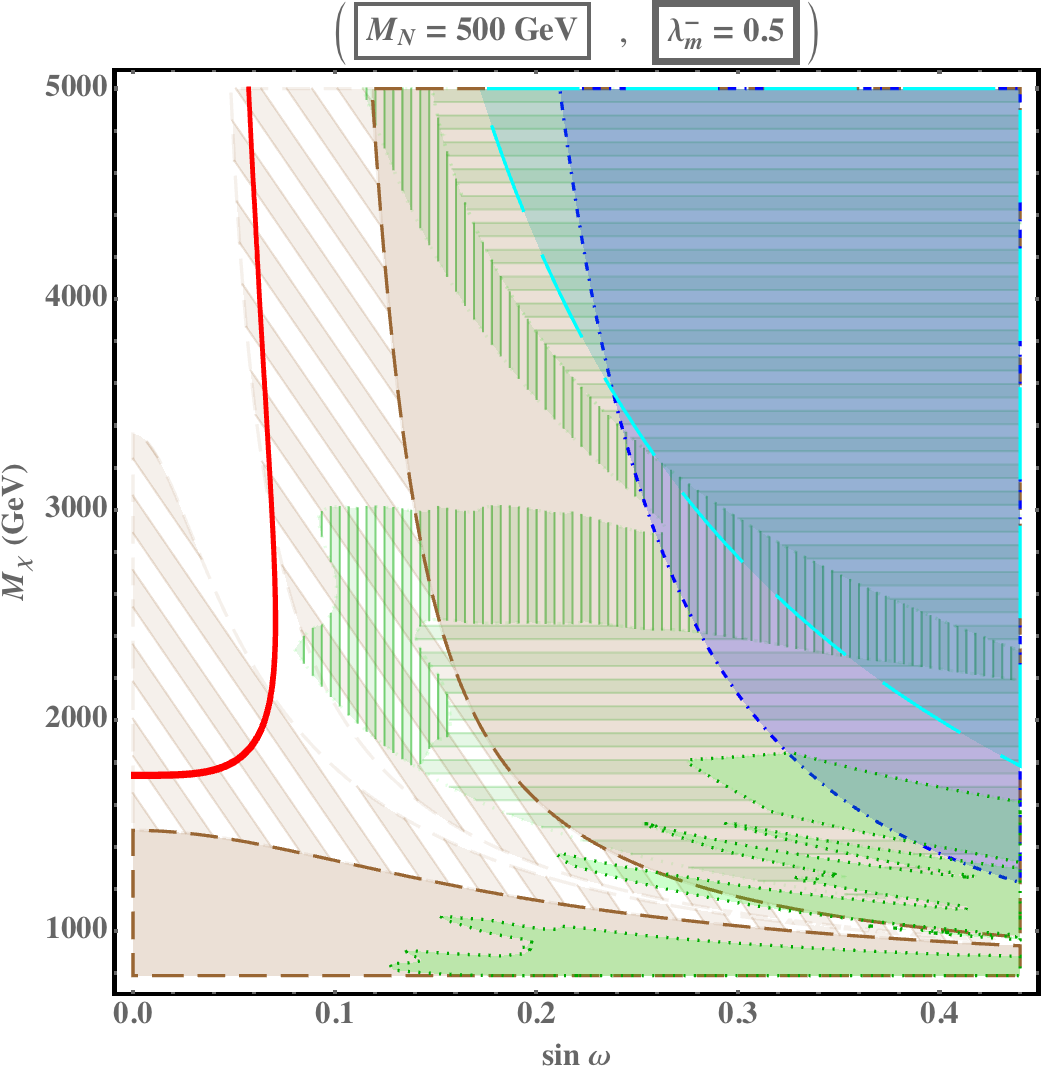}
\includegraphics[width=.329\textwidth]{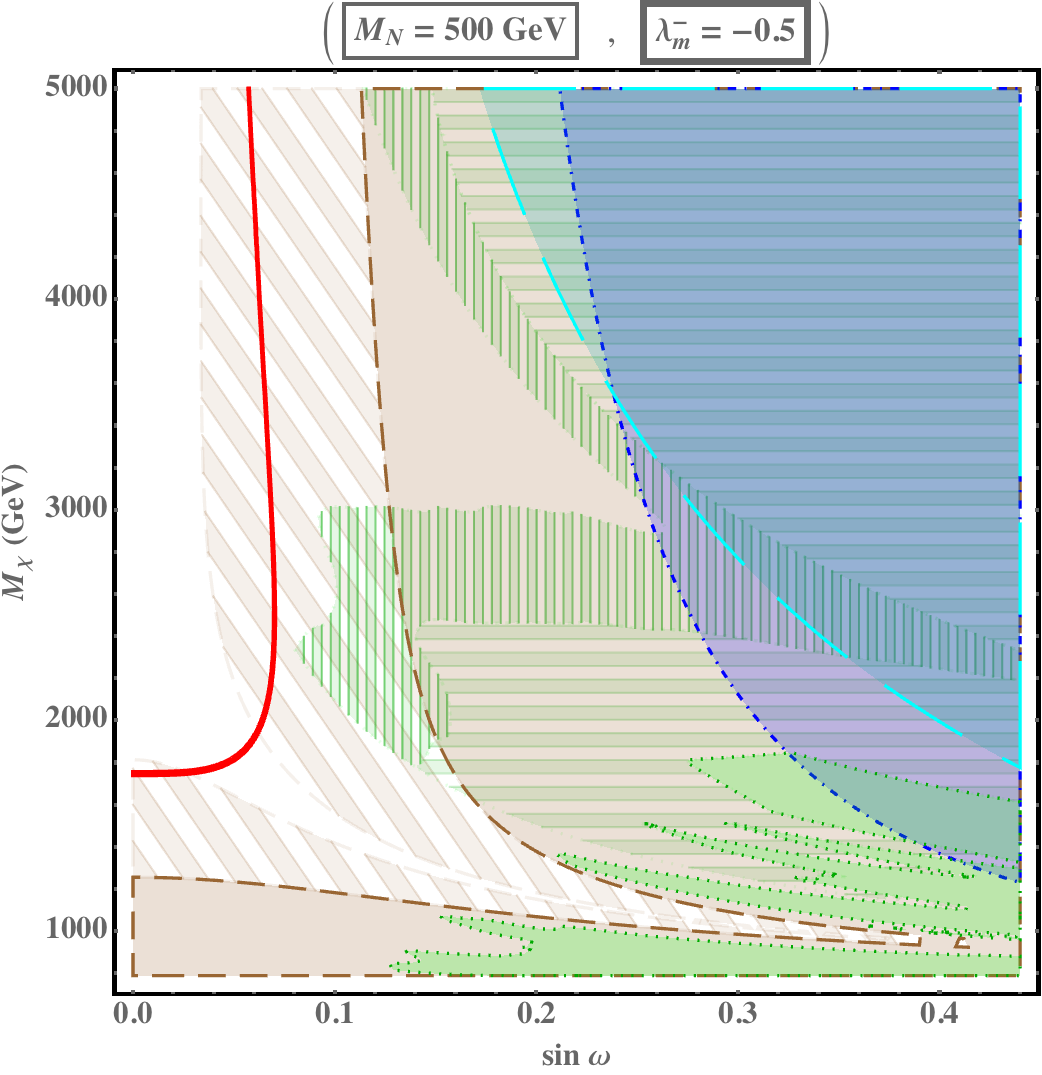}
\includegraphics[width=.329\textwidth]{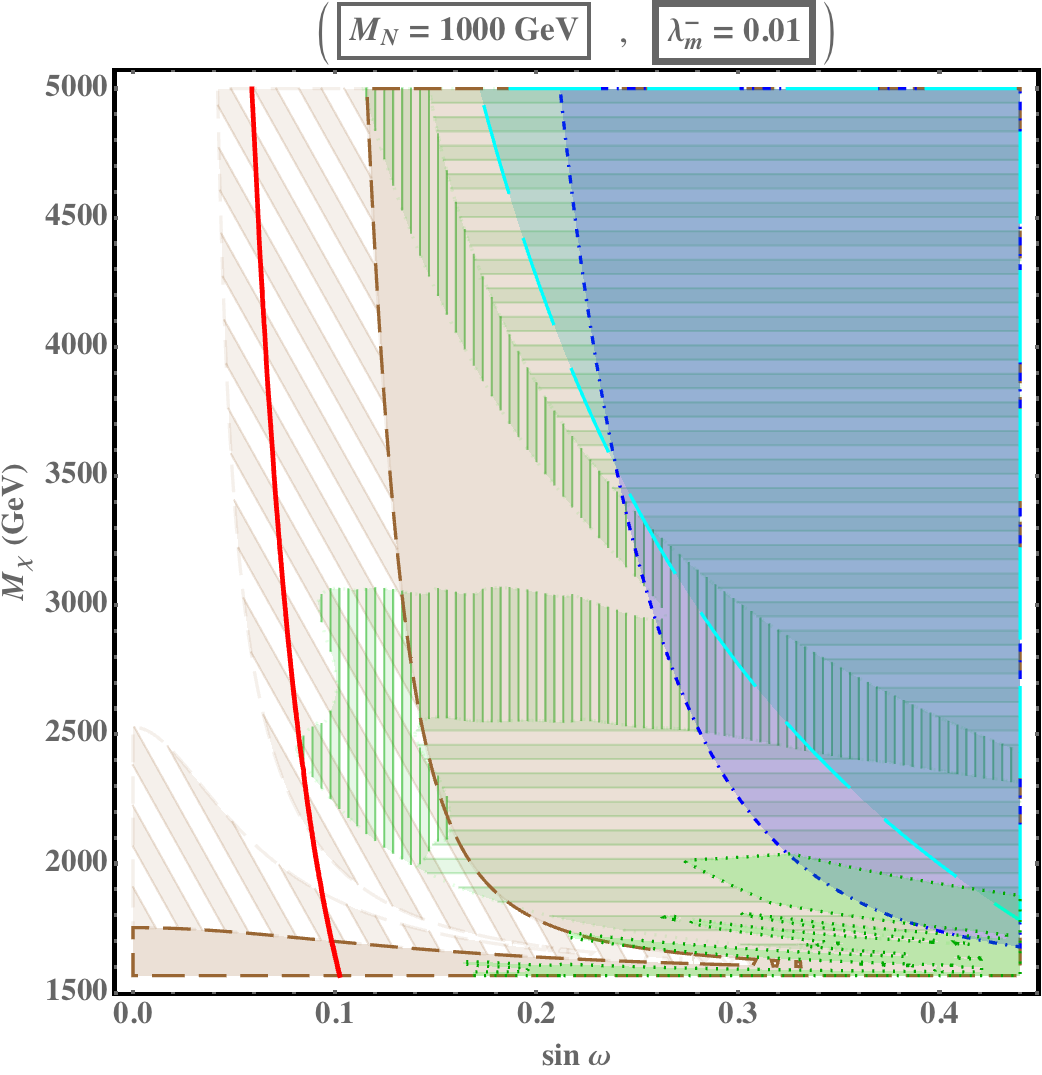}
\includegraphics[width=.329\textwidth]{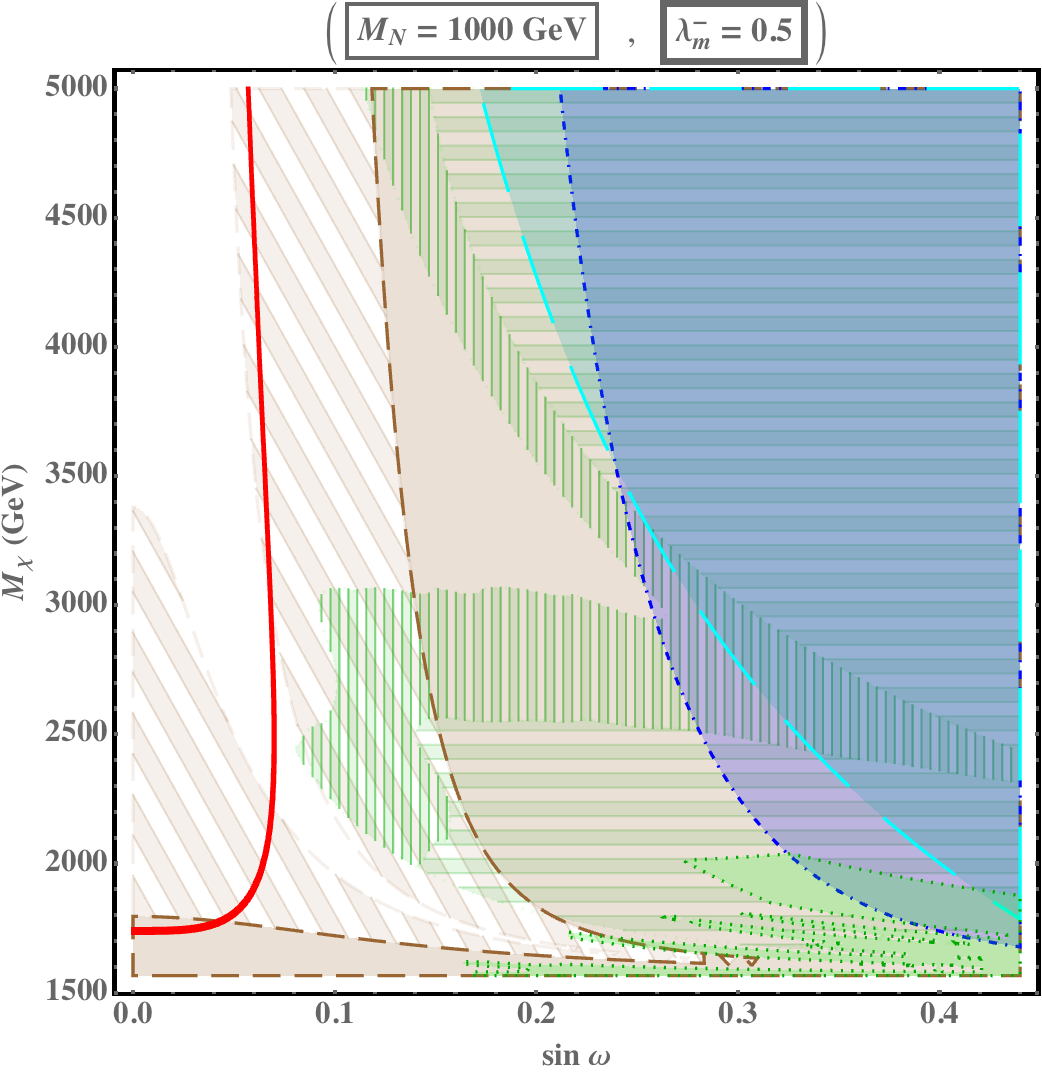}
\includegraphics[width=.329\textwidth]{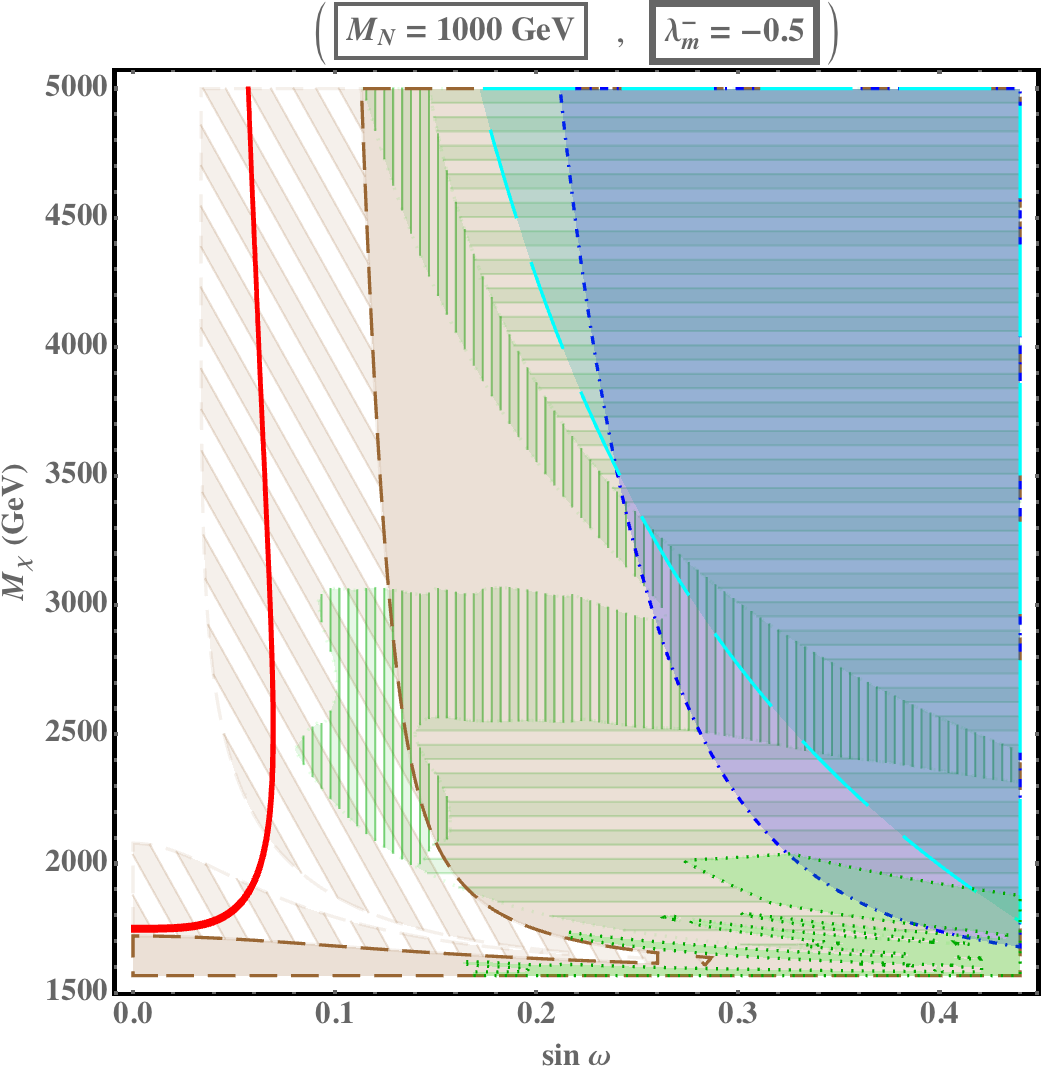}
\includegraphics[width=.329\textwidth]{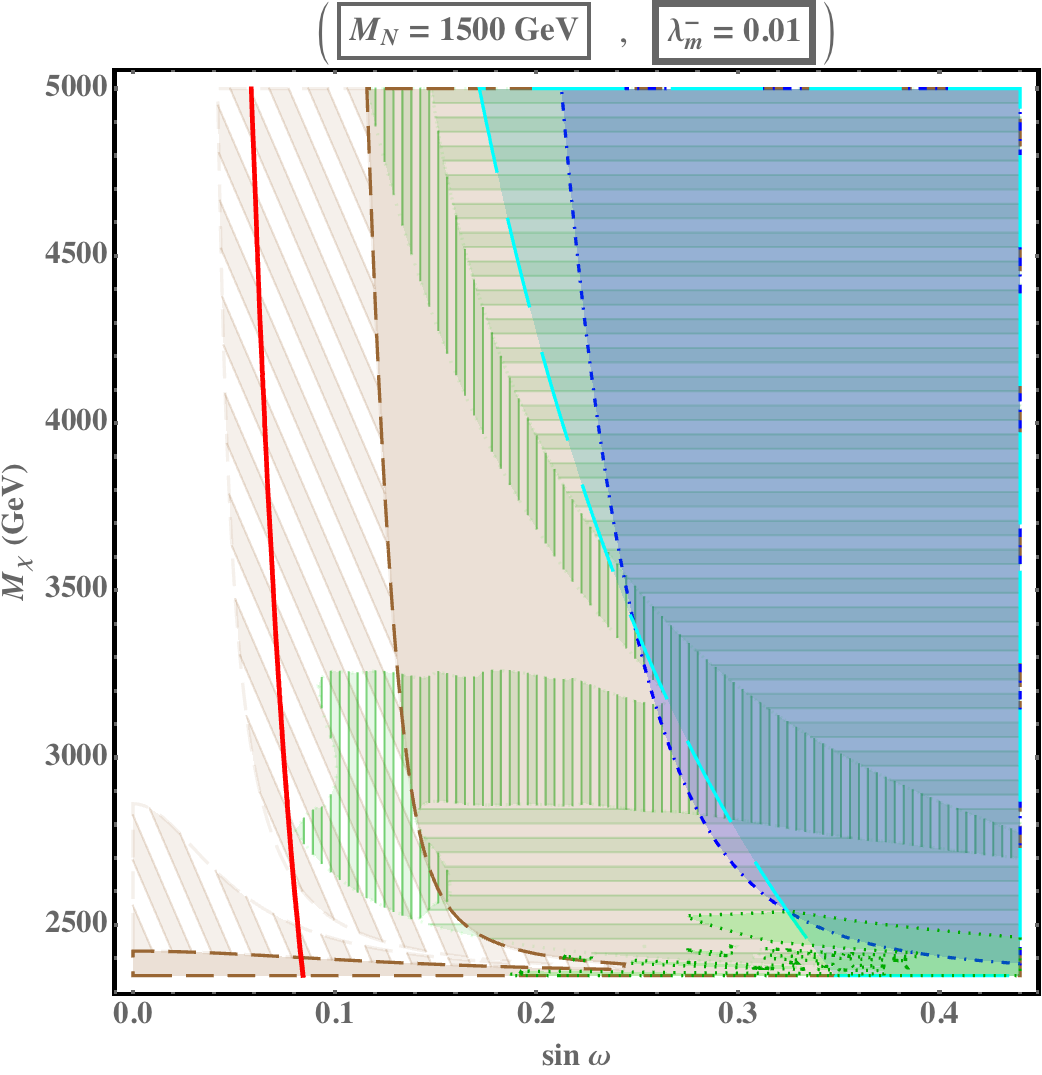}
\includegraphics[width=.329\textwidth]{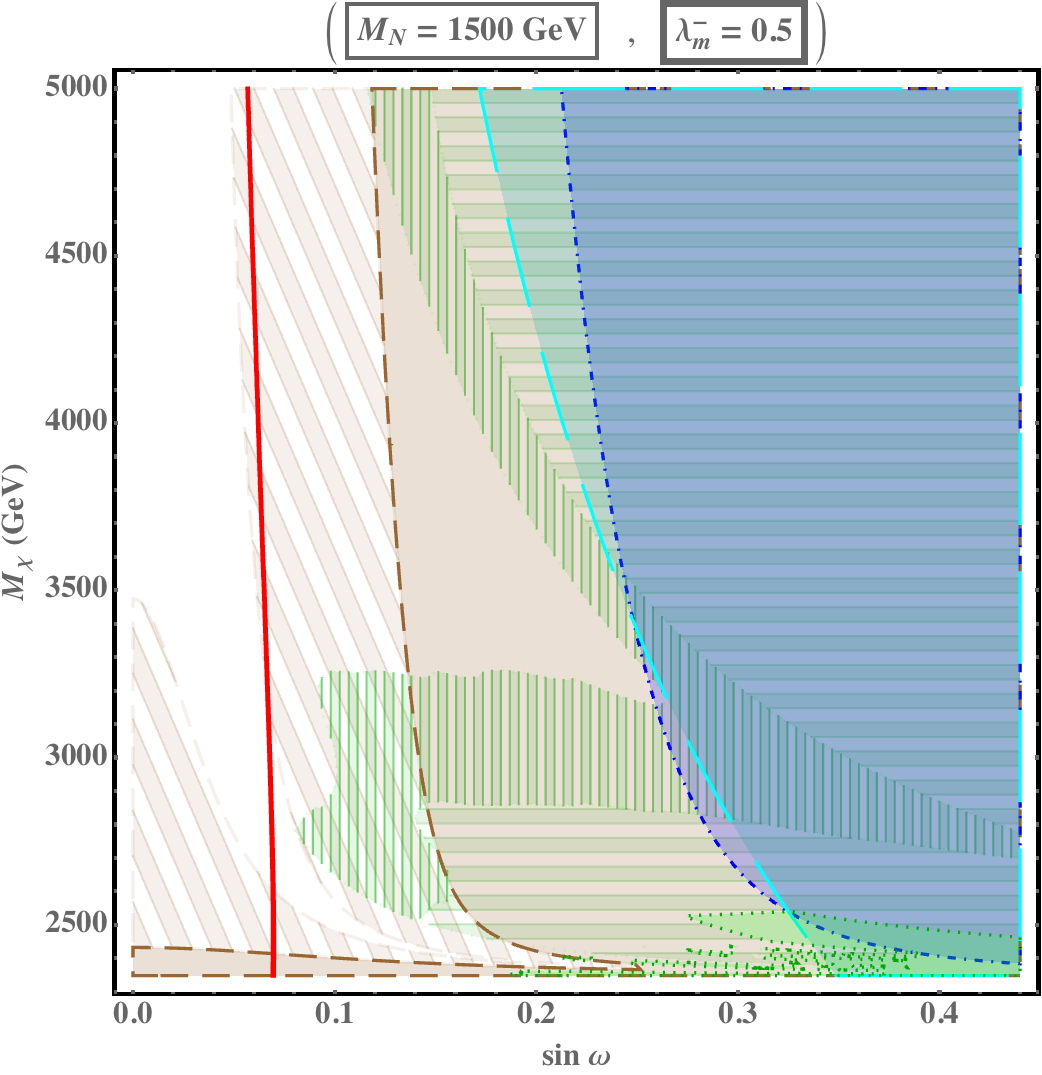}
\includegraphics[width=.329\textwidth]{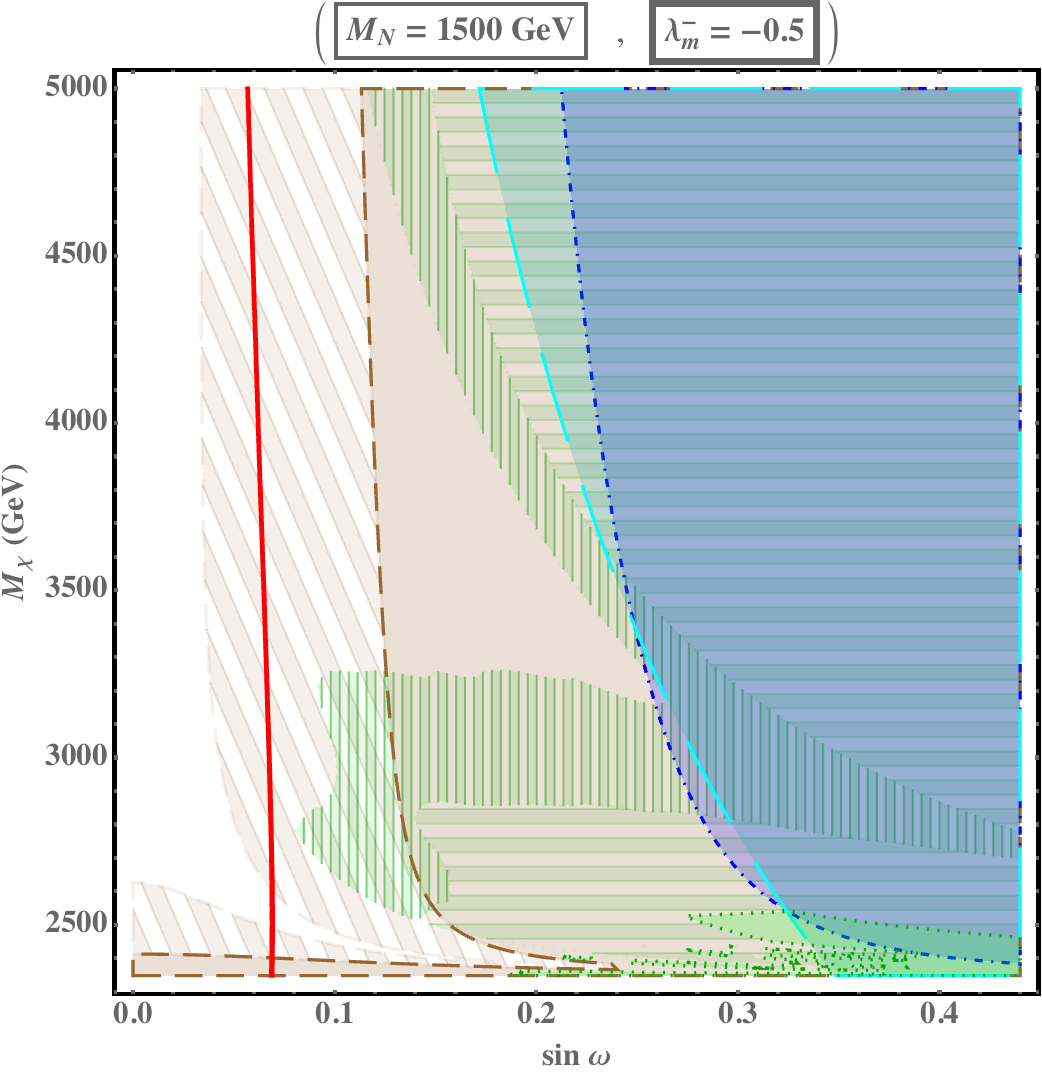}
\caption{Theoretical and experimental limits on the $\sin\omega-M_{\chi}$~plane, for benchmark values of the remaining two input parameters, $M_{N}$ and $\lambda_{m}^{-}$. All solid-colored regions are excluded, whereas the striped regions correspond to the projected future bounds by the LHC and the XENON1T experiments. (See the caption of Fig.~\ref{sinwms} for further details)}
\label{sinwMX}
\end{figure}

Moreover, the exclusion plots demonstrate that a lack of dark matter signal discovery by the XENON1T experiment further constrains the thermal relic abundance of a $\chi$~pseudoscalar, with a mass in the TeV~ballpark, as the dominant component of the dark matter in the Universe. This fact is most clearly exhibited in Fig.~\ref{sinwMX}, where the mass of the pseudoscalar WIMP, $M_{\chi}$, is plotted as a function of the mixing angle, $\sin\omega$. The upper limit of the vertical axis, $M_{\chi} \leq 5$~TeV, is based on the reported range of the results by the LUX direct detection experiment, whereas the lower limit is determined by the condition \eqref{staboneloop} for a given right-handed Majorana neutrino mass, $M_{N}$. The horizontal axis domain is the same as in Fig.~\ref{sinwms}, determined by the properties of the 125~GeV Higgs boson directly measured by the LHC. Considering similar benchmark values of the remaining two input parameters, $M_{N}$ and $\lambda_{m}^{-}$, as in Fig.~\ref{sinwms}, the aforementioned theoretical and experimental bounds and projections are incorporated within the plots.

The discussed observations and conclusions pertaining to the exclusion plots in Fig.~\ref{sinwms} remain also true in the parameter space depicted in Fig.~\ref{sinwMX}; upon a null discovery result, the ATLAS and XENON1T expected projections provide complementary constraints to the existing exclusion bounds, confining the amount of mixing between the two $CP$-even scalars to much smaller values, while a larger positive $\lambda_{m}^{-}$ further constrains lighter WIMP masses for small values of the mixing angle. In particular, one observes, within the large regions of the parameter space, that the XENON1T expected reach is able to probe the viability of the thermal relic abundance of a TeV-mass $\chi$~pseudoscalar WIMP forming an $\mathcal O (1)$ fraction of the dark matter in the Universe. We also note that the narrow opening window within the LUX direct detection constraints, as well as within the XENON1T projected limits, is attributed to the destructive interference between the $t$-channel scalar exchanges, as explained below \eqref{DMdirCS}.

\begin{figure}
\includegraphics[width=.329\textwidth]{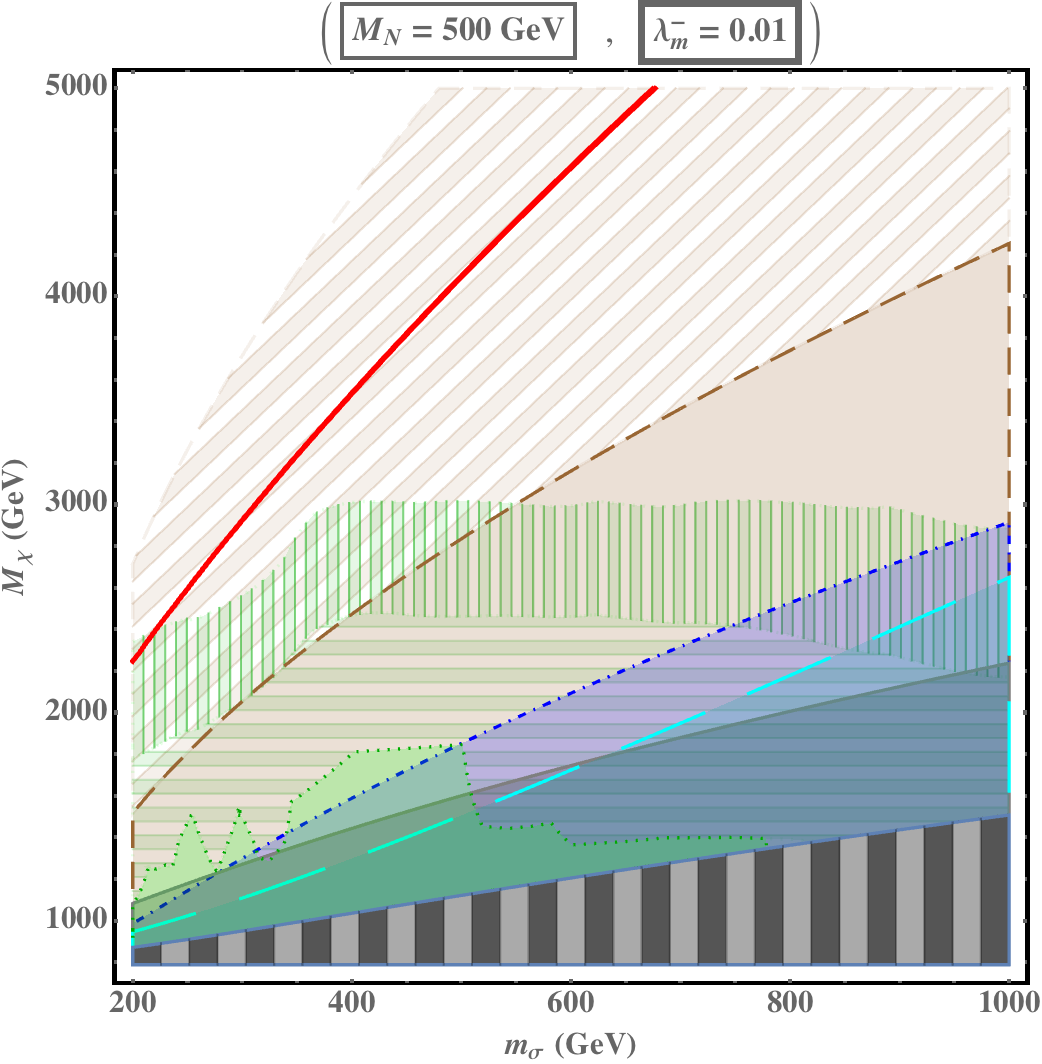}
\includegraphics[width=.329\textwidth]{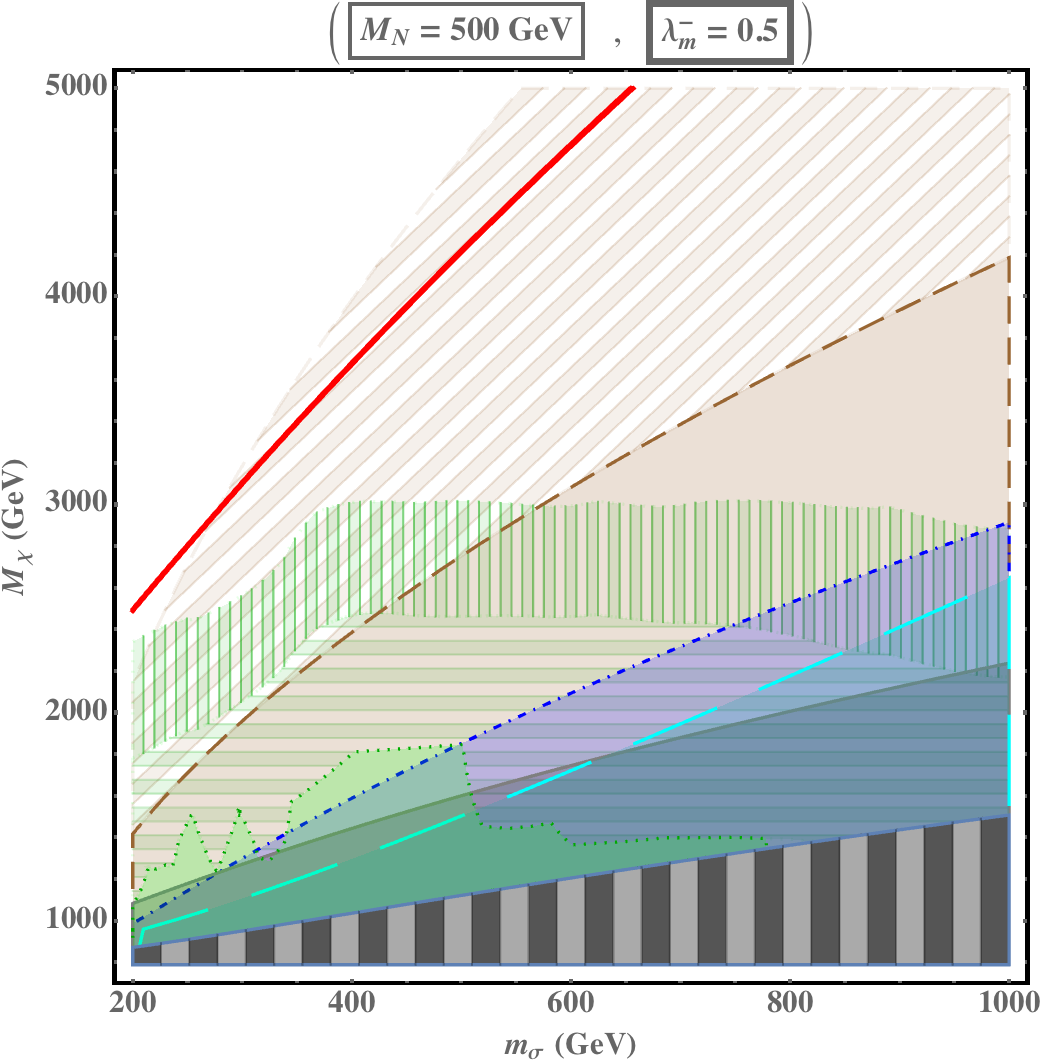}
\includegraphics[width=.329\textwidth]{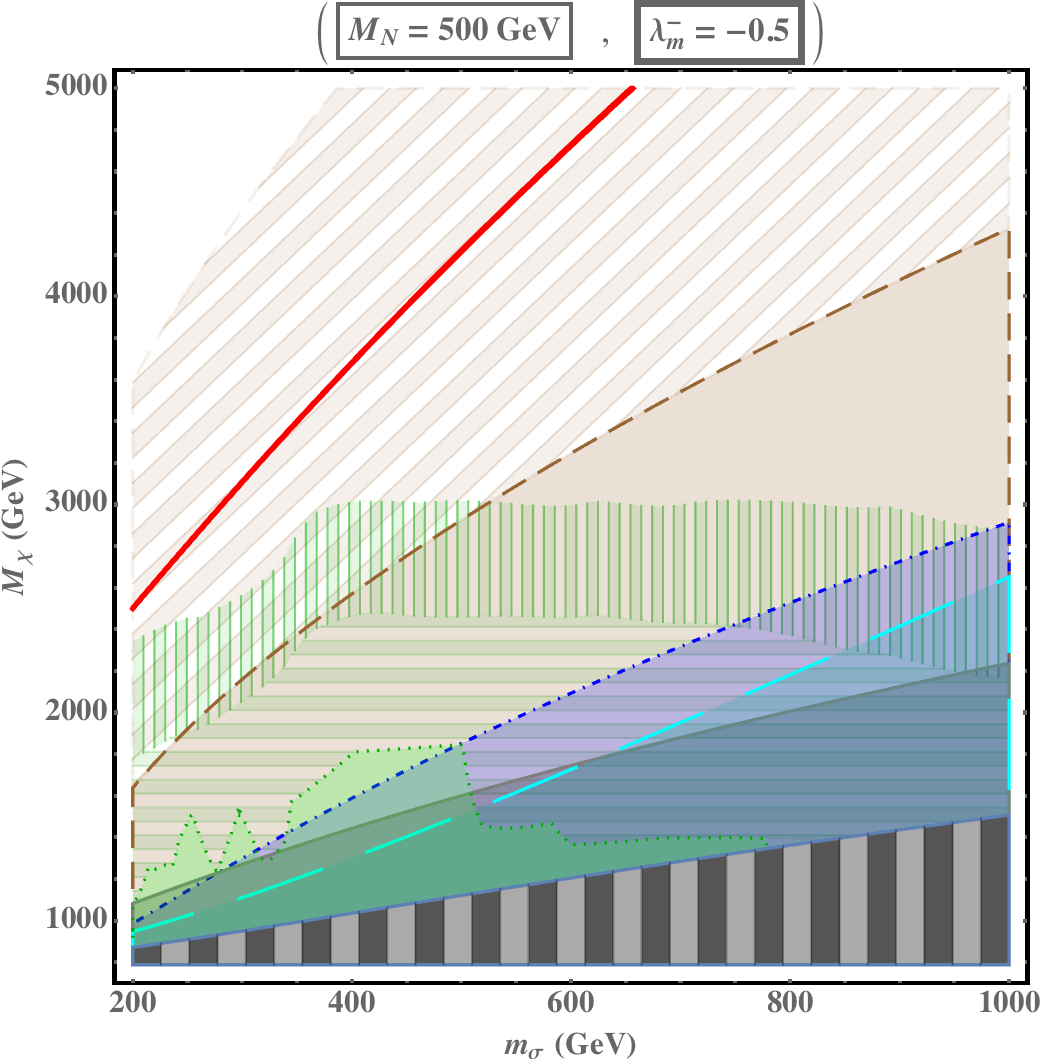}
\includegraphics[width=.329\textwidth]{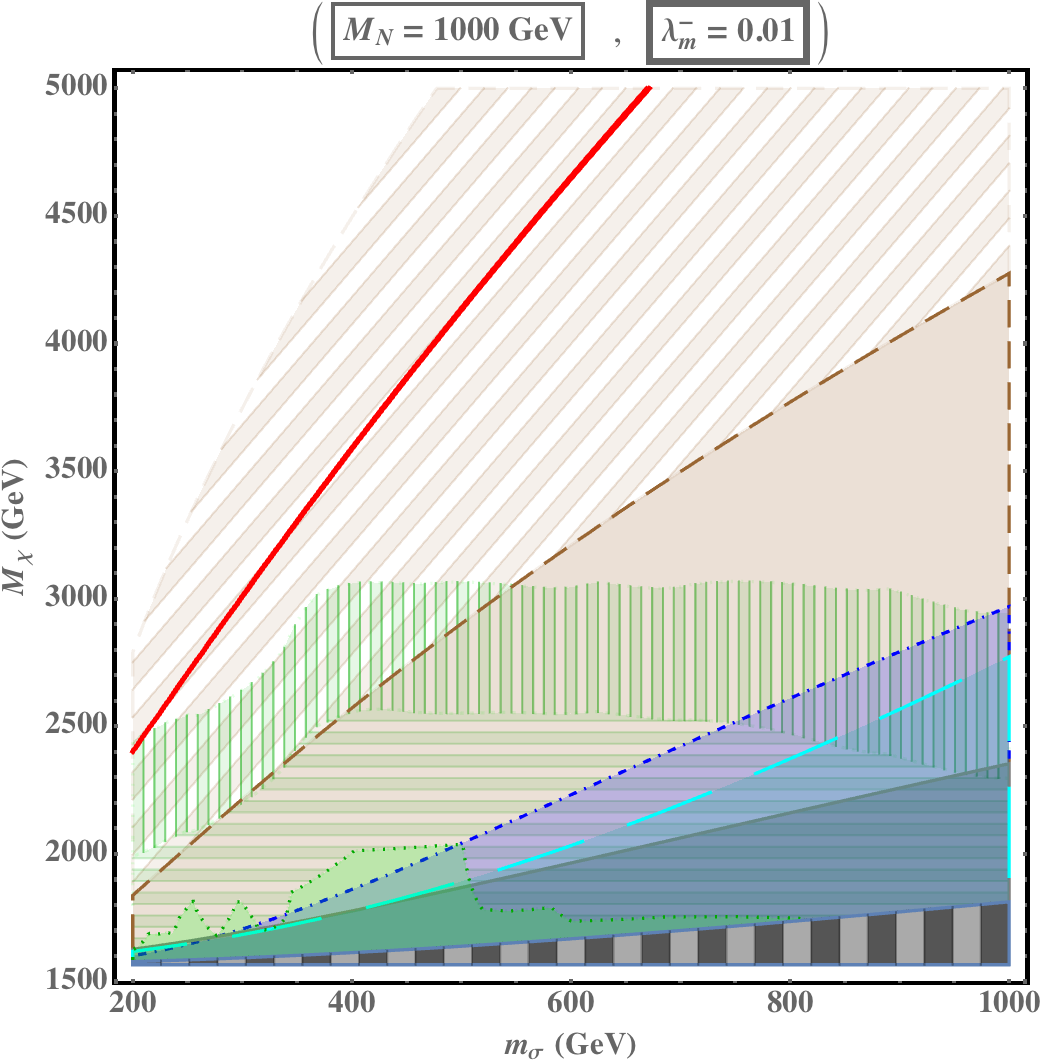}
\includegraphics[width=.329\textwidth]{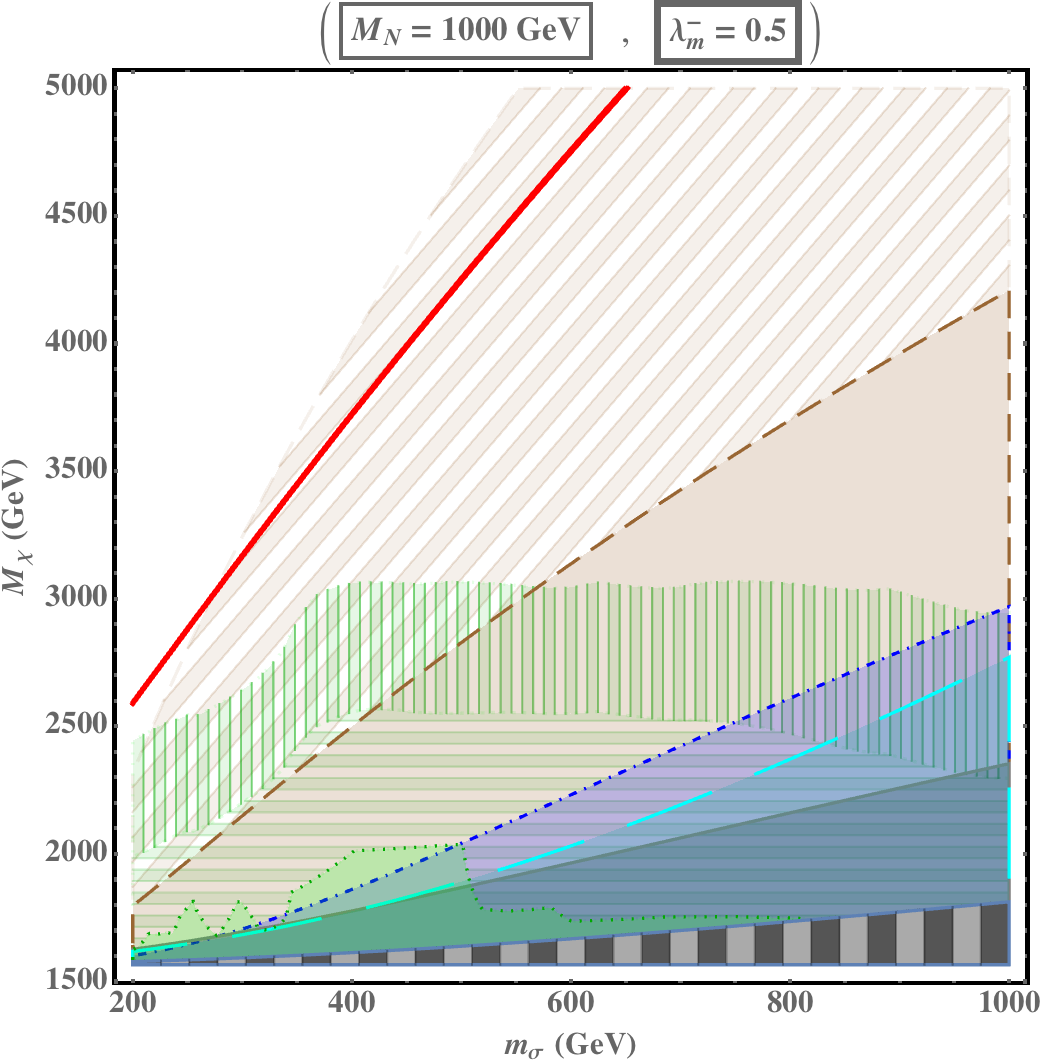}
\includegraphics[width=.329\textwidth]{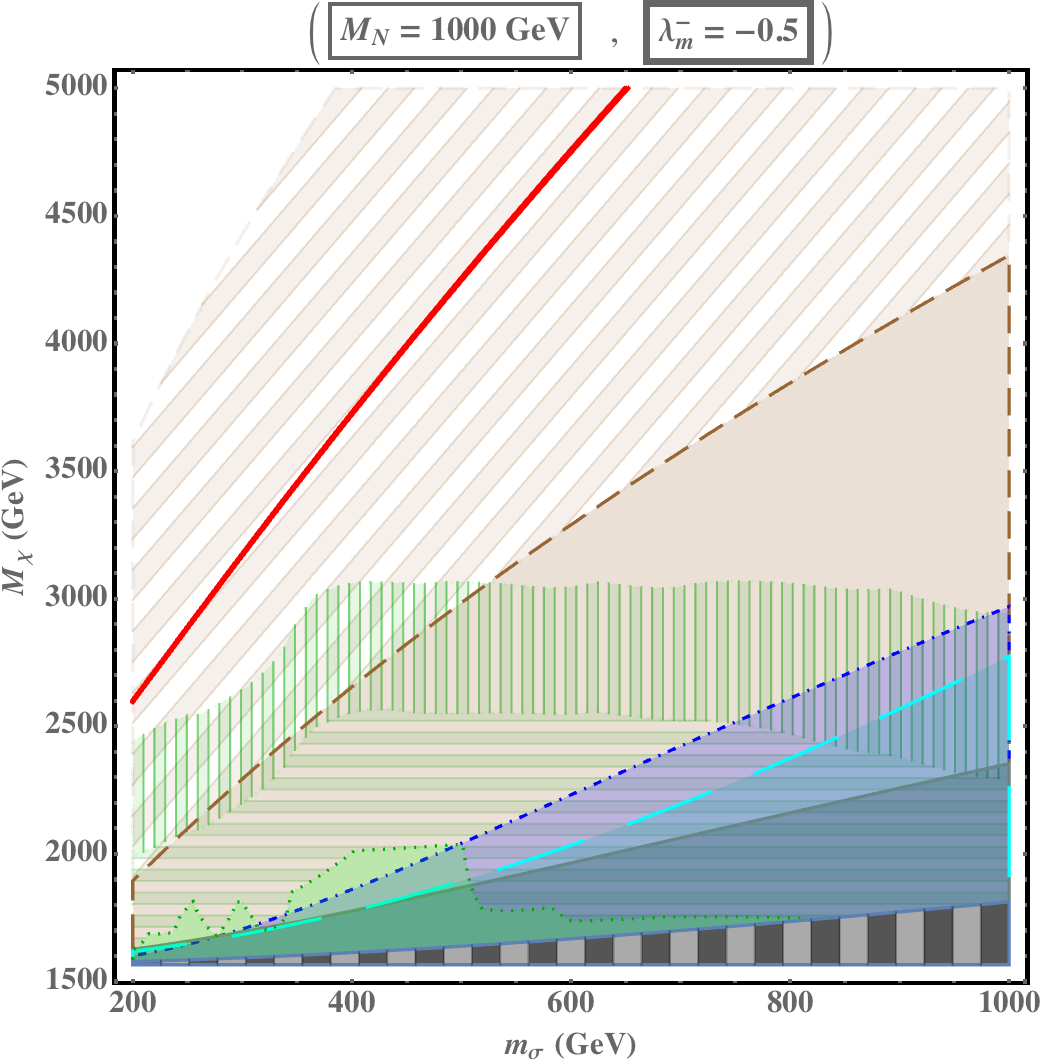}
\includegraphics[width=.329\textwidth]{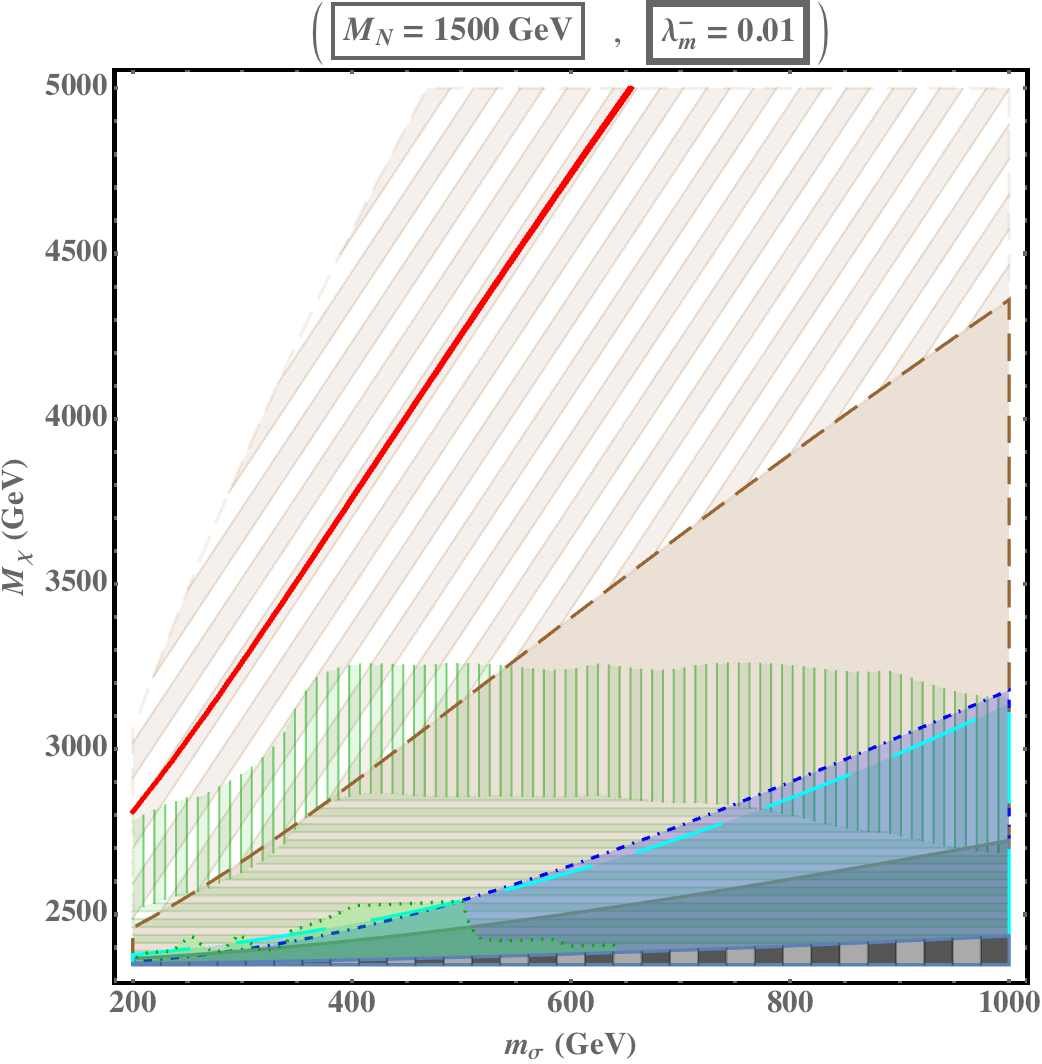}
\includegraphics[width=.329\textwidth]{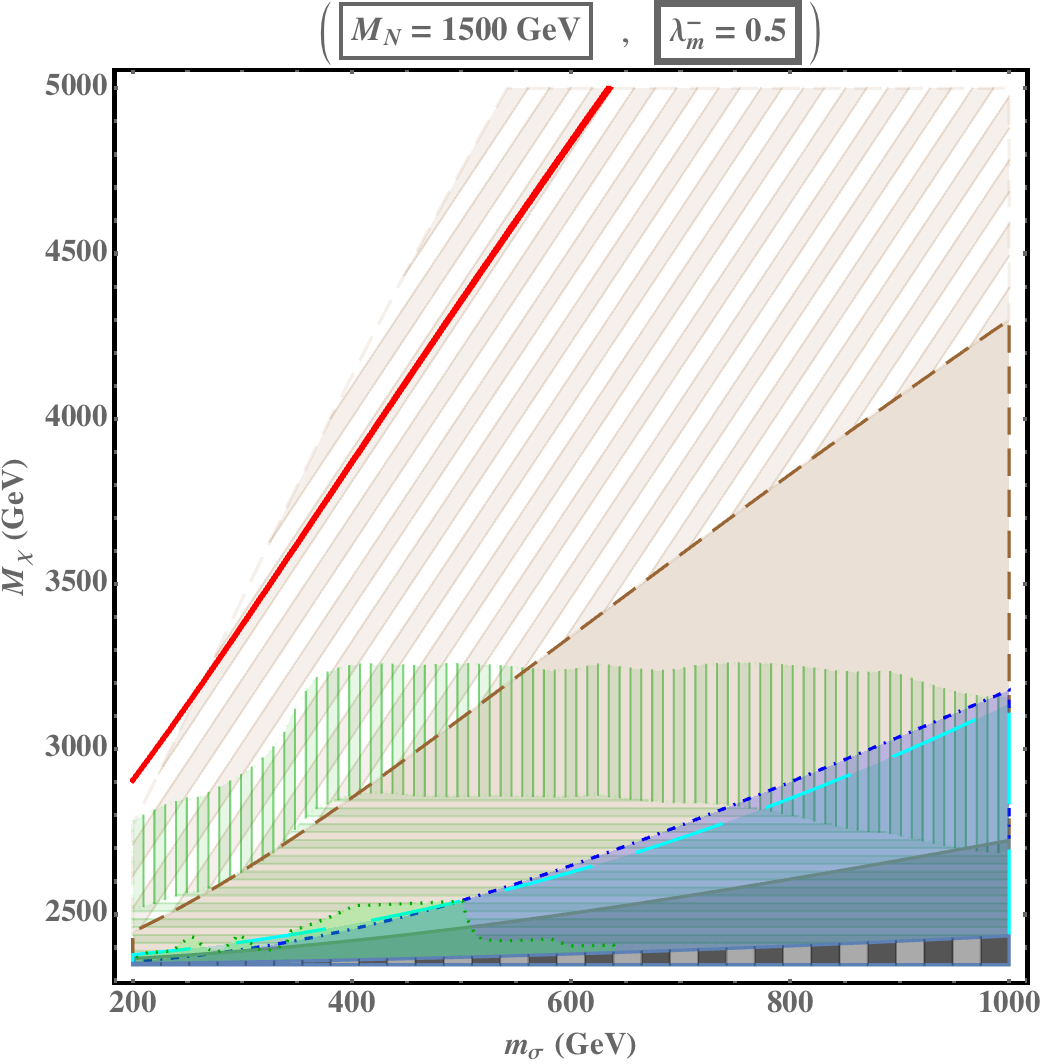}
\includegraphics[width=.329\textwidth]{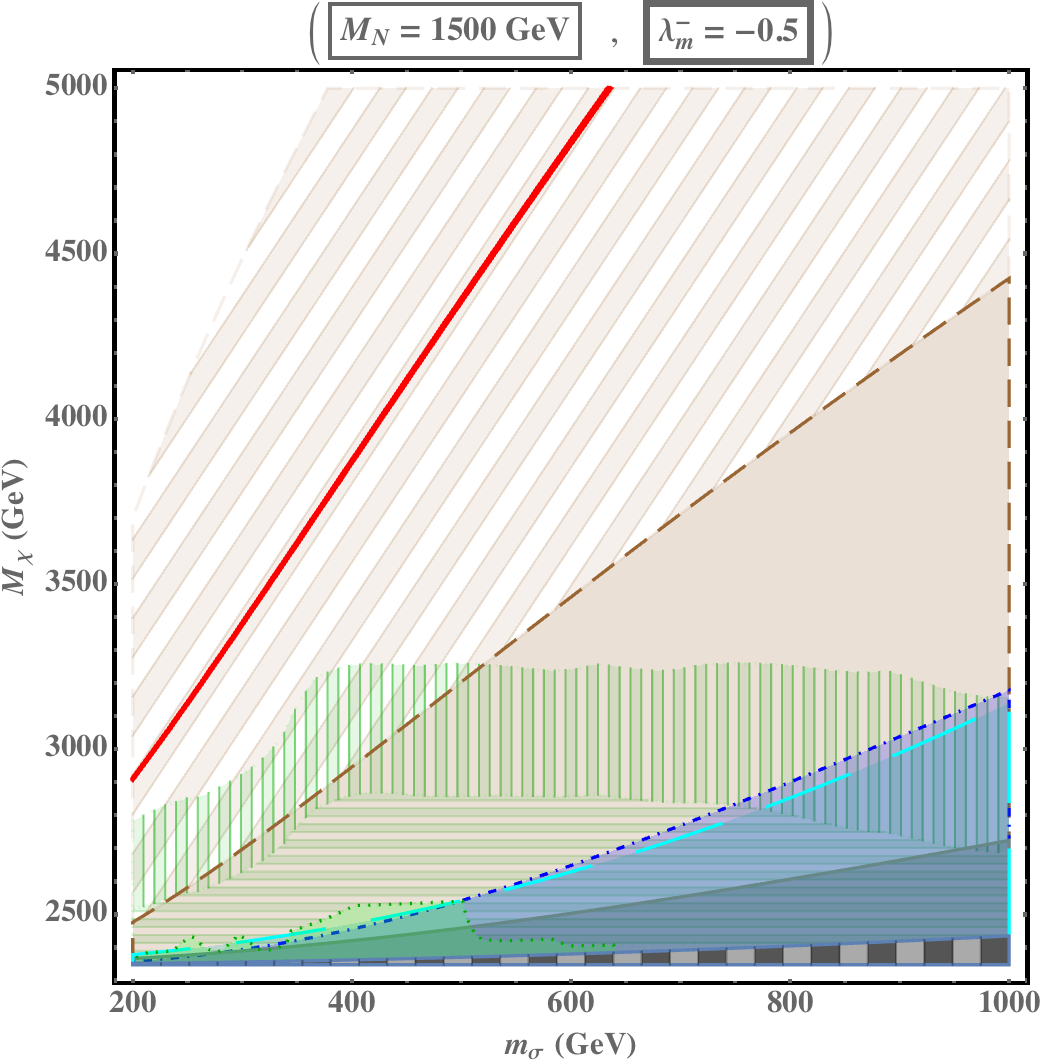}
\caption{Theoretical and experimental limits on the $m_{\sigma}-M_{\chi}$~plane, for benchmark values of the remaining two input parameters, $M_{N}$ and $\lambda_{m}^{-}$. All solid-colored regions are excluded. The thick black-gray vertically striped region at the bottom of the plots is mathematically forbidden by the $|\sin\omega| \leq1$ requirement, whereas the solid gray region above it is excluded due to $\sin\omega \lesssim 0.44$ at 95\%~C.L. by the LHC measurements of the 125~GeV Higgs properties. All other striped regions correspond to the projected future bounds by the LHC and the XENON1T experiments. (See the caption of Fig.~\ref{sinwms} for further details)}
\label{msMX}
\end{figure}

A direct interplay between the mass of the $\sigma$~scalar and the $\chi$~pseudoscalar is presented in Fig.~\ref{msMX}, for the discussed ranges of the axes, and similar choices of the remaining two free parameters, $M_{N}$ and $\lambda_{m}^{-}$, as in the previous exclusion plots. Superimposing the theoretical and the experimental bounds and projections within the $m_{\sigma}-M_{\chi}$~plane demonstrates, once more, that the expected ATLAS and XENON1T discovery prospect are able to adequately constrain large regions of the parameter space of the model, if no heavy Higgs-like scalar or dark matter signal is detected by these experiments. Specifically, the viable region of the depicted parameter space is ``cornered'' toward heavier $\chi$~pseudoscalar dark matter and lighter $\sigma$~scalar masses by the XENON1T projected limits, which, as in the previous figures, challenge the thermal relic abundance curve of a TeV-mass $\chi$~pseudoscalar constituting the dominant dark matter component.

\section{Stability of the Potential and the Triviality Condition}\label{TrivStab}

In this section, we devote attention to analyzing the stability of the scalar potential \eqref{V0} and the triviality requirement. These conditions were previously studied in \cite{Farzinnia:2013pga} with a cutoff at 100~TeV; here, we extend the analysis consistently up to the Planck scale. Specifically, we demand that the conditions \eqref{stabtree} hold for the running couplings as a function of the renormalization scale, $\mu$ (not to be confused with \eqref{mupar}). This guarantees, to the leading order, the vacuum stability of the renormalization group (RG)-improved scalar potential until the Planck energy is reached. Furthermore, we demand that none of the running couplings encounters a Landau pole within the considered energy regime; the latter requirement is specifically materialized by selecting $4\pi$ as the upper bound for the magnitude of the running couplings.

The SM gauge, the top Yukawa, and the low energy Higgs quartic couplings are fixed at their $\overline{\text{MS}}$-scheme determined values \cite{Zoller:2014cka} at the top mass, $\mu = m_t$,
\begin{equation}\label{SMrun}
g(m_{t}) = 0.6483\,, \quad g'(m_{t}) = \sqrt{3/5}\times 0.3587\, , \quad g_{c}(m_{t}) = 1.1671\, , \quad y_{t}(m_{t}) = 0.9369\, , \quad\lambda_{\phi}(m_{t}) = 6\times0.1272\ .
\end{equation}
We normalize the hypercharge coupling according to $g' = \sqrt{3/5} \, g_{1}$, where $g_{1}$ denotes the corresponding coupling with the GUT normalization. Moreover, we employ a different normalization for $\lambda_{\phi}$ in \eqref{V0} with respect to the conventional normalization in the literature by a factor of 6, as reflected in \eqref{SMrun}. 

As discussed in Sec.~\ref{rev}, the flat direction of the potential is defined at the energy scale $\Lambda$, given by \eqref{Lambfin}, which is a function of the dark matter mass, $M_{\chi}$, and the mass of the right-handed Majorana neutrinos, $M_{N}$. For the current analysis, we fix the values of the six high energy scalar quartic couplings at the minimization energy scale, $\Lambda$, according to the expressions \eqref{paraIV2}. The latter determine four of these couplings in terms of the free parameters of the theory \eqref{inputs}, noting that the $\lambda_{m}^{-}$ and $\lambda_{\chi}$ couplings are free parameters on their own at this energy scale. The right-handed neutrinos' Yukawa coupling, $y_{N}$, is fixed according to the corresponding expression in \eqref{paraIV2} at the energy scale $\mu = M_{N}$.

The running behavior of the couplings as a function of the renormalization scale, $\mu$, is determined by solving the RG~equation, $d \mathcal C/d \log \mu = \beta_{\mathcal C}$, where $\mathcal C$ denotes any of the running couplings, and $\beta_{\mathcal C}$ its corresponding $\beta$-function. The one-loop $\beta$-functions for all of the model's couplings have been computed in \cite{Farzinnia:2013pga}, and summarized in the Appendix~\ref{betafun} for convenience. As evident from \eqref{SMgauge} and \eqref{Yukawa}, the RG~equations for each of the gauge and the Yukawa couplings can be solved independently, whereas those for the scalar quartic couplings \eqref{scalar} form a coupled system and must be solved simultaneously. Imposing the vacuum stability conditions \eqref{stabtree} for the running couplings up to the Planck scale, as well as demanding no Landau poles to be present, results in additional bounds upon the free parameter space of the model.

\begin{figure}
\includegraphics[width=.4\textwidth]{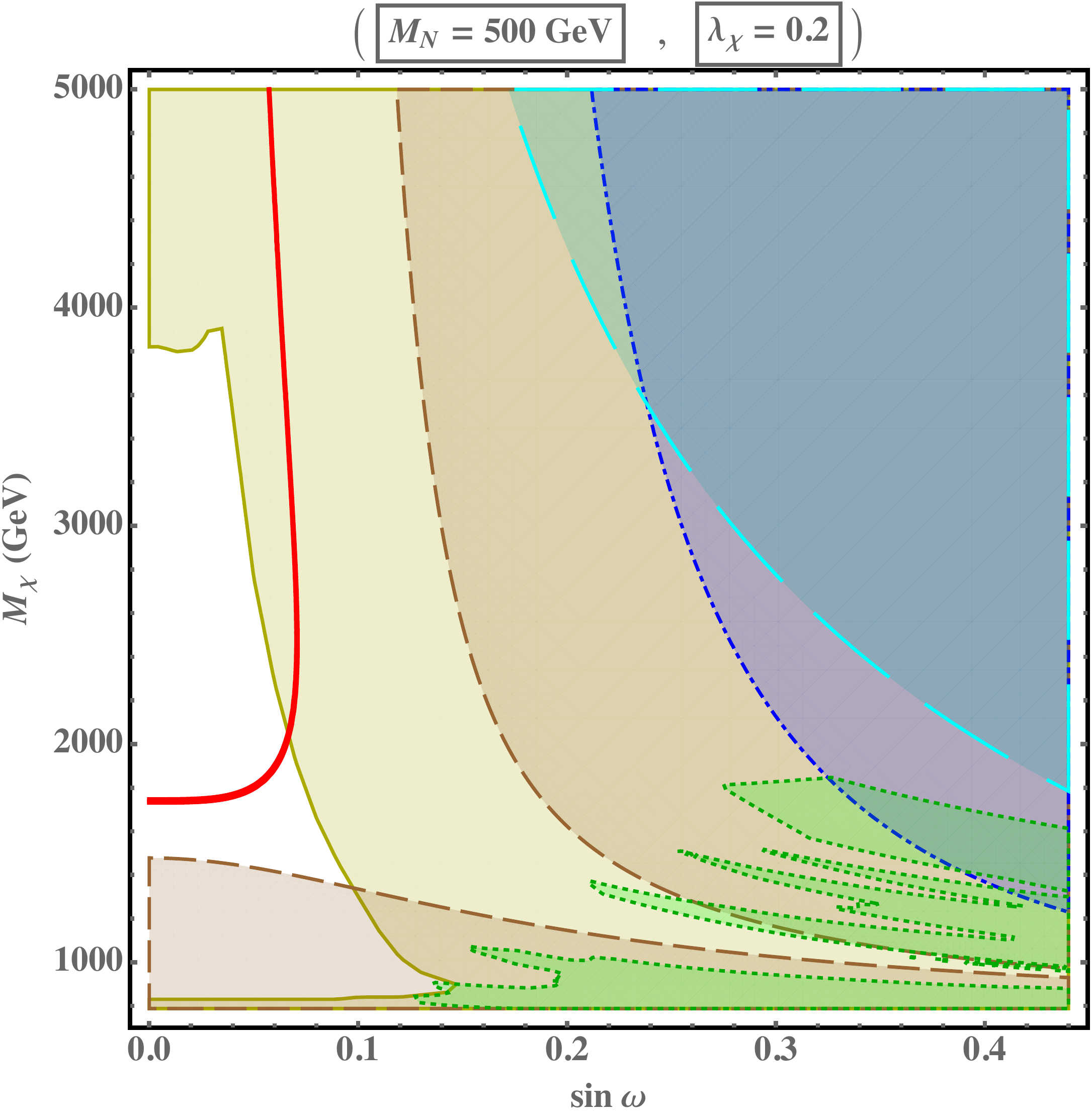}
\includegraphics[width=.4\textwidth]{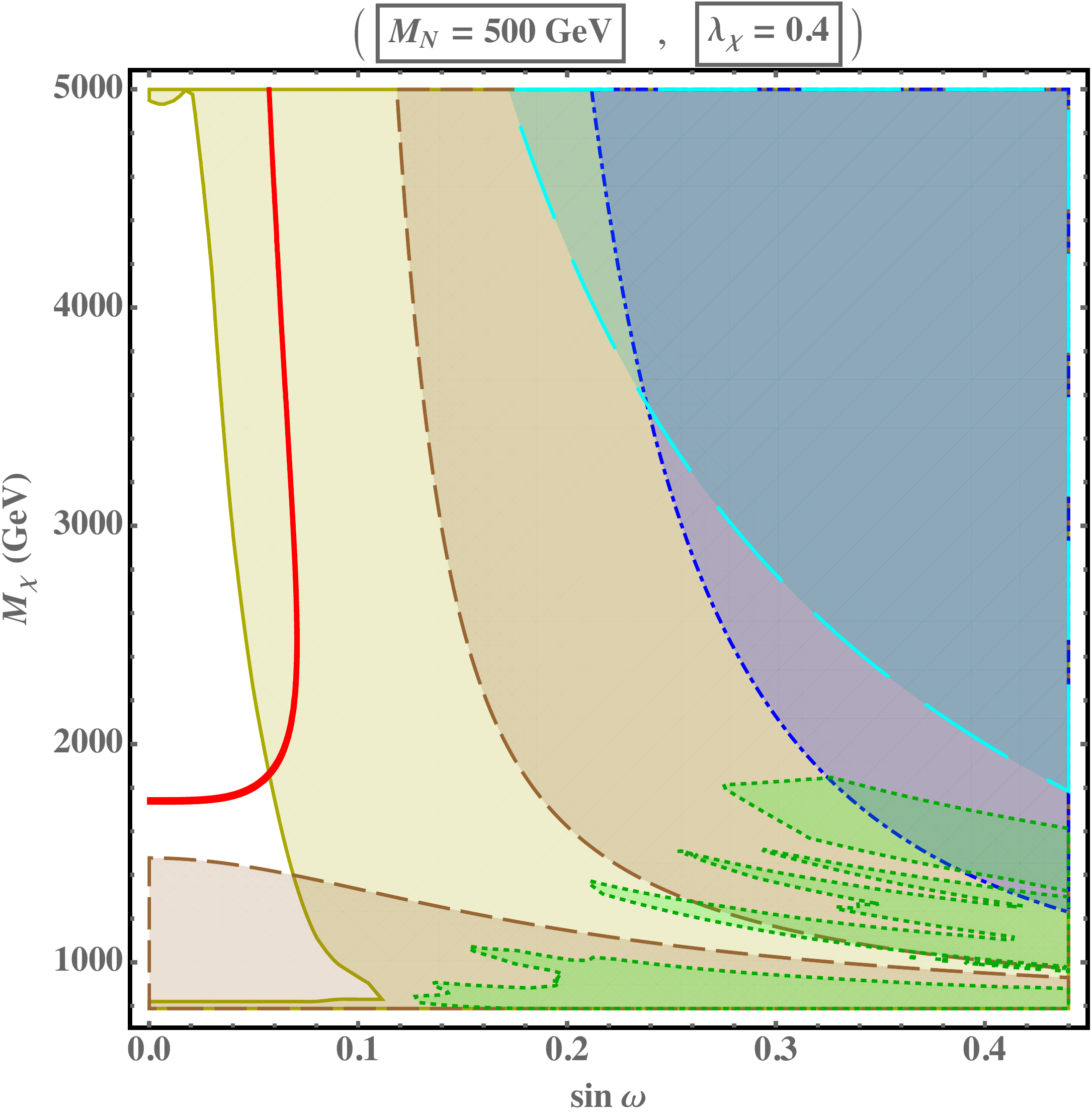}
\includegraphics[width=.4\textwidth]{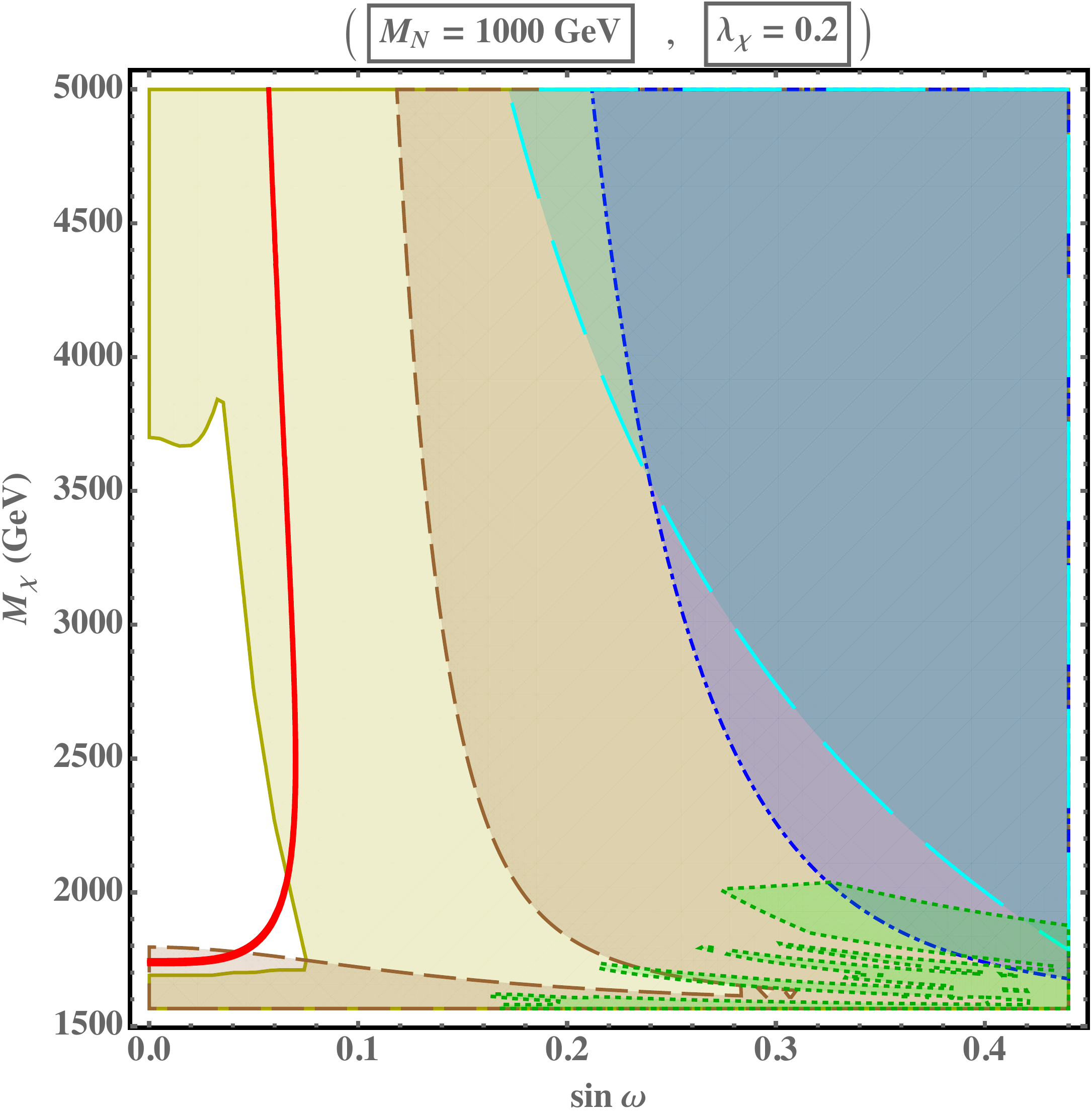}
\includegraphics[width=.4\textwidth]{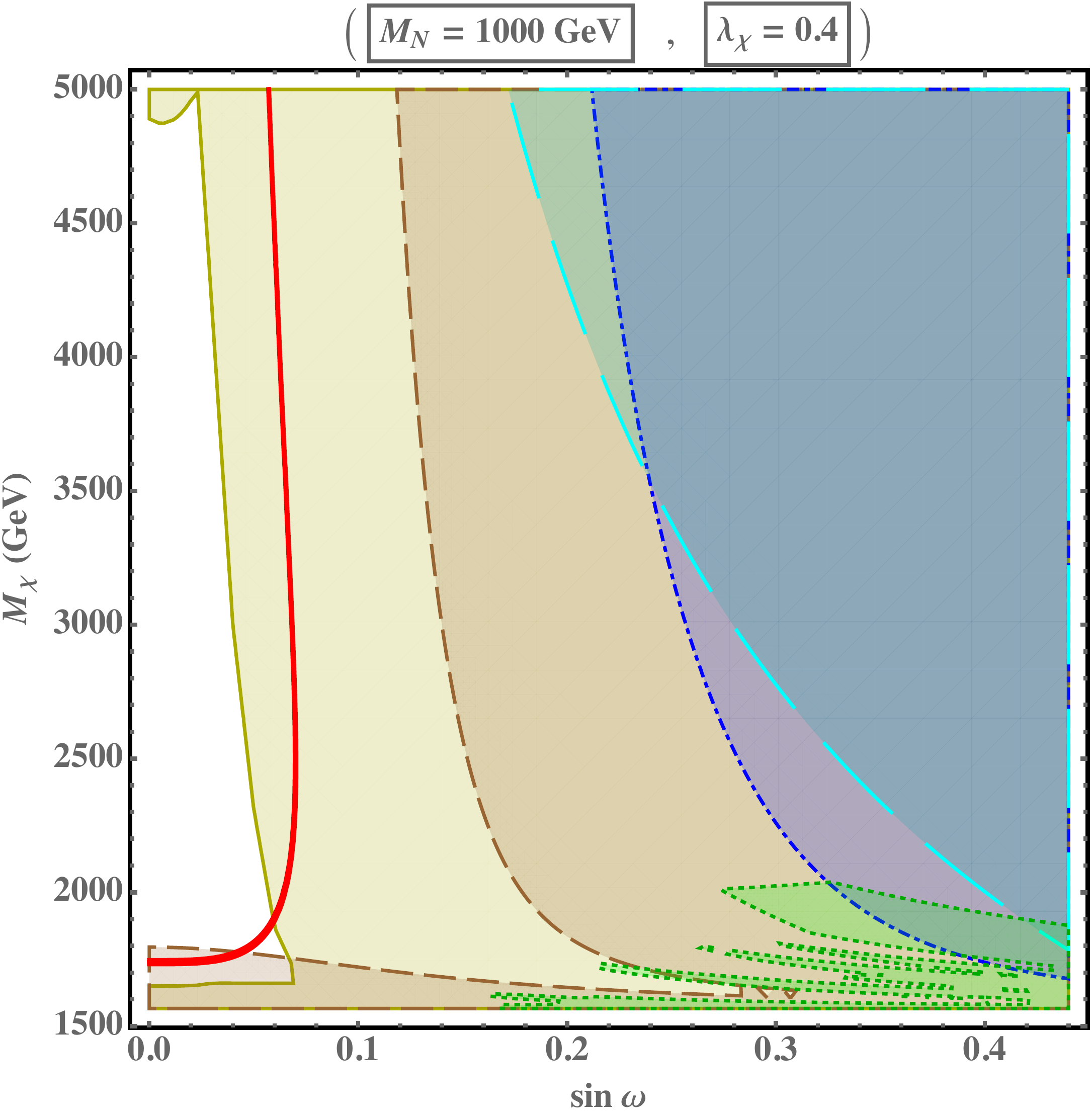}
\caption{Theoretical and experimental limits on the $\sin\omega-M_{\chi}$~plane, for benchmark values of the input parameters, $M_{N}$ and $\lambda_{\chi}$. The value of the remaining free parameter has been fixed at $\lambda_{m}^{-} = +0.5$ in all cases. The stability and triviality bounds (solid yellow region) have been incorporated within the exclusion plots together with the other discussed constraints. The upper bound on the dark matter mass, $M_{\chi}$, for smaller values of the mixing angle, is due to requiring the stability of the potential up to the Planck scale, whereas the subsequent sharply declining upper bound curve, at larger mixing angles, signifies the $\lambda_{\chi}$ developing a Landau pole before reaching that scale. The additional lower bound on $M_{\chi}$, slightly more stringent than its theoretical lower limit \eqref{staboneloop}, is attributed to the matching scale, $\Lambda$ (c.f. \eqref{Lambfin}), diverging near this formal lower limit, once more destabilizing the effective potential. All solid-colored regions are excluded. (See the caption of Fig.~\ref{sinwms} for further details)}
\label{sinwMXStabTriv}
\end{figure}

The as such determined vacuum stability and triviality constraints are presented in Fig.~\ref{sinwMXStabTriv}, where, the dark matter mass is plotted vs the mixing angle. These $\sin\omega - M_{\chi}$~panels incorporate, in addition, the previously discussed theoretical and experimental bounds arising from the perturbative unitarity, the electroweak precision tests, the LHC heavy Higgs searches, the LUX dark matter direct detection results, as well as the WIMP thermal relic abundance reported by the Planck collaboration.\footnote{For the sake clarity and accessibility of the figures, we do not include the projected bounds, discussed in Sec.~\ref{prosp}, in the current figures.} The value of the $\lambda_{m}^{-}$ coupling is universally set at +0.5 in all panels; while, several benchmarks of the remaining two free parameters, $\lambda_{\chi}$ and $M_{N}$, are considered.

As evident from Fig.~\ref{sinwMXStabTriv}, the stability and triviality bounds separate three distinct regions within each panel. First, at the smaller values of the mixing angle, an upper bound on the dark matter mass is imposed by the vacuum stability condition. This is attributed to the fact that, for a given $M_{N}$, a larger $M_{\chi}$ results in a larger value of the minimization scale $\Lambda$, where the initial conditions for the high energy scalar couplings were defined. As a consequence, the positive bosonic contribution from $\lambda_{m}^{-}$ to the running of the Higgs quartic coupling~$\lambda_{\phi}$ (c.f. \eqref{scalar}) is ``postponed'' to a larger energy scale, which will then be inadequate to compensate for the negative fermionic top Yukawa contribution. Hence, the $\lambda_{\phi}$~coupling turns negative before reaching the Planck scale.\footnote{The Higgs quartic coupling within the ordinary SM is known to turn negative at $\mu \sim 10^{10}$~GeV for the experimentally measured central values of the Higgs boson and the top quark mass \cite{Degrassi:2012ry}.} Furthermore, one notes that, although a positive initial value of $\lambda_{m}^{-}$ provides a positive contribution to the $\lambda_{\phi}$~running, the $\lambda_{m}^{+}$~coupling always starts negative (c.f. \eqref{flatdir} and \eqref{lmp}), contributing to the destabilization of the effective potential. Therefore, the competing effects between these two couplings (and their entangled runnings) within the $\lambda_{\phi}$~$\beta$-function results in the nontrivial shape of the upper bound ``plateau'' at small mixing angle values. This constraint is only mildly dependent on the right-handed neutrino mass, $M_{N}$, which enters into the analysis primarily via the positive $y_{N}$~Yukawa contributions to the running of the $\lambda_{m}^{+}$~coupling. Since the latter starts negative, invoking in a destabilizing effect on the potential, a larger $M_{N}$ mildly aggravates the bound. In addition, one can deduce from \eqref{Lambfin} that a heavier right-handed Majorana neutrino leads to a larger minimization scale $\Lambda$ for a given dark matter mass, which also leads to a slightly more stringent bound. In contrast, a larger value of the $\lambda_{\chi}$~coupling enhances the running of $\lambda_{m}^{-}$ and its subsequent positive contribution to the $\lambda_{\phi}$~running. This in turn allows for larger dark matter masses before the vacuum stability condition is violated; hence, mitigating this constraint.

The second region is defined by the upper $M_{\chi}$ bound for (slightly) larger values of the mixing angle, and is characterized by the sharply declining curve as a function of $\sin\omega$. This constraint arises due to the triviality condition, as $\lambda_{\chi}$ develops a Landau pole below the Planck scale, in the region above and to the right of the curve. Larger values of $M_{\chi}$ and $\sin\omega$ necessitate a larger $\lambda_{\eta\chi}$ by \eqref{paraIV2}, leading to a swift running of the $\lambda_{\chi}$~coupling, and hastening the development of a Landau pole. This triviality bound is also mildly sensitive to $M_{N}$, which affects the running of $\lambda_{\eta\chi}$ via the positive $y_{N}$~Yukawa contributions. Therefore, a larger $M_{N}$, once more, mildly aggravates the bound. A greater value of $\lambda_{\chi}$, on the other hand, results in a more pronounced constraining effect, given its direct influence on its own running.

As the third region, one identifies a lower bound on the dark matter mass, slightly more stringent than its theoretical lower limit, which is mildly dependent on the mixing angle values. This constraint also arises due to the vacuum stability condition. Specifically, a dark matter mass near its formal lower limit \eqref{staboneloop} corresponds to an almost vanishing $\beta$~coefficient \eqref{beta} within the one-loop effective potential. This, in turn, causes the minimization scale, $\Lambda$, to diverge (c.f. \eqref{Lambfin}), leading to a destabilization of the effective potential in analogy with the dark matter mass upper bound within the first region, as elaborated above. Hence, a very heavy dark matter as well as a dark matter mass near the formal lower limit result in a too large minimization scale, $\Lambda$, subsequently destabilizing the potential. The sensitivity of this dark matter mass lower bound to the $M_{N}$ and $\lambda_{\chi}$ parameters is also completely analogous to the upper bound's case within the first region, and exhibits a similar behavior.

Finally, we comment on the sign and the magnitude of the free parameter $\lambda_{m}^{-}$, in light of the vacuum stability and triviality study, which has been fixed at +0.5 in this treatment. Our performed analysis implies that $\lambda_{m}^{-}$ cannot substantially deviate from this value; a larger selected $\lambda_{m}^{-}$ quickly drives the $\lambda_{\chi}$~coupling toward a Landau pole before reaching the Planck scale, whereas a smaller value of $\lambda_{m}^{-}$ cannot adequately compensate for the negative contributions introduced by $\lambda_{m}^{+}$ and $y_{t}$ within the $\beta$-function of the Higgs quartic coupling $\lambda_{\phi}$. As a result, the latter runs into negative values below the Planck scale, violating the stability of the vacuum. This effect occurs more severely if the initial value of $\lambda_{m}^{-}$ is chosen to be negative, in spite of being formally allowed at the tree-level (c.f.~\eqref{stabtree}). In conclusion, given the sensitivity of the vacuum stability and triviality analysis on the $\lambda_{m}^{-}$~parameter, a (substantial) deviation of its value from +0.5 severely reduces the viable region of the parameter space.

It is evident from the panels in Fig.~\ref{sinwMXStabTriv} that demanding the vacuum stability of the effective potential up to a Planck scale cutoff confines the dark matter mass to the TeV~ballpark, whereas the condition of the absence of any Landau poles up to that same cutoff favors a small mixing between the SM Higgs boson and the singlet scalar, $\sin\omega \lesssim 0.06$. These constraints are complementary to the previously derived theoretical and experimental bounds, further narrowing the viable region of the parameter space and increasing the predictive power of the model. In particular, in light of these results, the thermal relic abundance of the $\chi$~pseudoscalar as the main component of the WIMP dark matter in the Universe strongly favors sub-TeV right-handed Majorana neutrino (degenerate) masses, since heavier right-handed neutrinos start to impose substantial tension on this assumption.

\section{Conclusion}\label{concl}

The current treatment has been devoted to studying the impacts of the future experimental constraints, expected at the LHC and the XENON1T dark matter detector, as well as the formal bounds arising from the vacuum stability and triviality considerations up to the Planck scale, on the input parameters of the minimal classically scale invariant and $CP$-symmetric extension of the SM. In addition to the SM content, the scenario includes one complex gauge singlet scalar and three (mass-degenerate) flavors of the right-handed Majorana neutrinos; the latter facilitate the see-saw mechanism and induce masses for the SM neutrinos. The gauge singlet pseudoscalar component, protected by the $CP$-symmetry of the potential, forms a viable WIMP dark matter candidate, while the gauge singlet scalar component, mixed with the SM Higgs boson, is identified in the mass basis as the pseudo-Nambu-Goldstone boson of the (approximate) scale symmetry. All the mass scales are dynamically induced via the Coleman-Weinberg mechanism at the quantum level, and the formalism introduces five new input parameters.

We applied the ATLAS prospects for discovering a heavy SM-like Higgs scalar, at $\sqrt s = 14$~TeV with an integrated luminosity of 300 and 3000~fb$^{-1}$ within the mass range 200-1000~GeV, to the pseudo-Nambu-Goldstone boson of the scenario and demonstrated that the expected projections will considerably constrain the free parameter space of the model, if no such heavy Higgs is discovered. In addition, a lack of dark matter signal discovery at the forthcoming XENON1T direct detection experiment implies further complementary constraints on the model's parameters. Demonstrating our results in extensive exclusion plots covering various benchmarks of the input parameters, we conclude that these future experiments are capable of probing vast regions of the model's parameter space; specifically, the combined discovery prospects for the collider and dark matter direct detection experiments confine a mixing between the SM Higgs boson and the singlet scalar to $\sin\omega \lesssim 0.04$ in most cases, while exploring the possibility of the TeV~scale $\chi$~pseudoscalar and its thermal relic abundance constituting the dominant dark matter component in the Universe. To facilitate the comparison of the additional impact of the projected data on the parameter space with the existing bounds, various previously analyzed formal and experimental limits \cite{Farzinnia:2013pga,Farzinnia:2014xia} have also been superimposed within all exclusion plots.

\begin{figure}
\includegraphics[width=.4\textwidth]{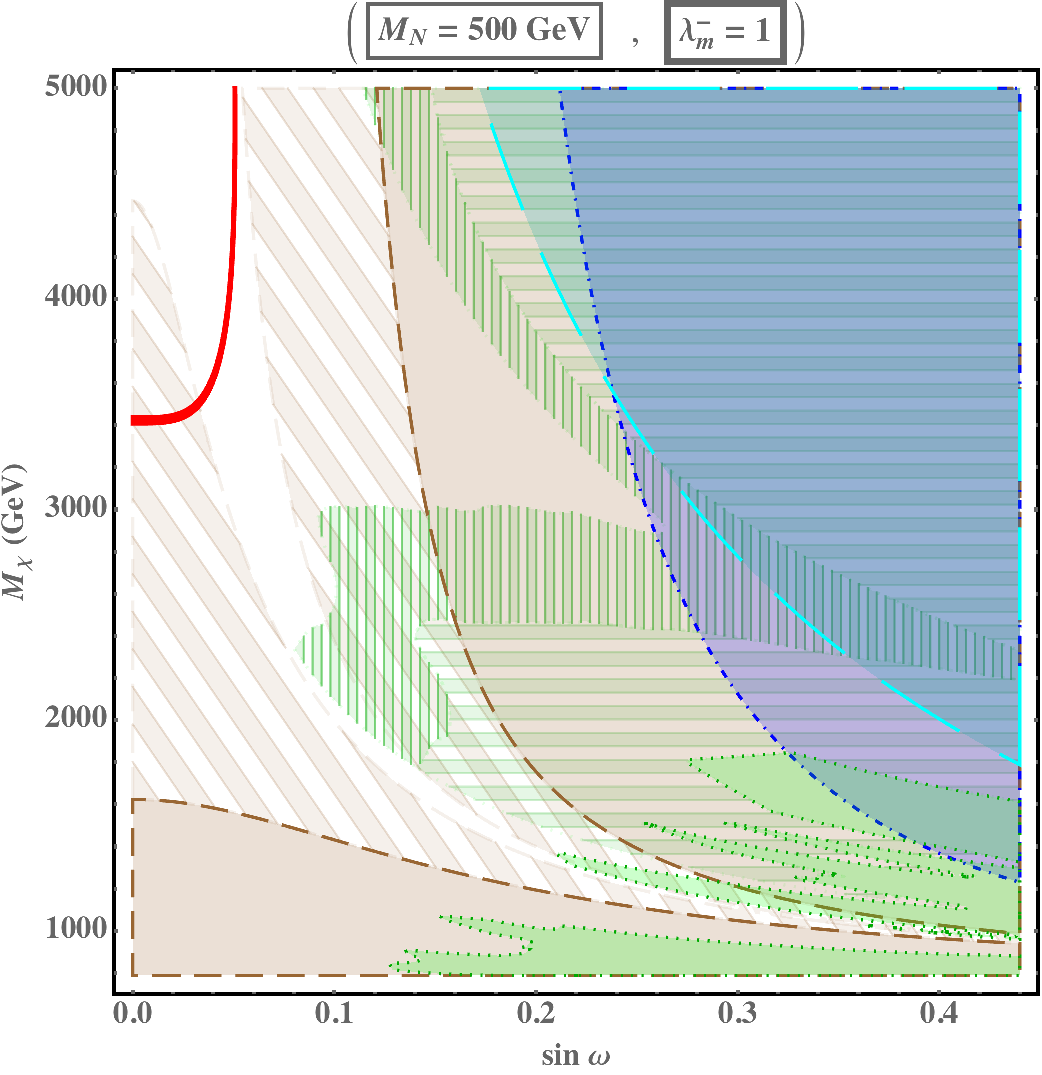}
\includegraphics[width=.4\textwidth]{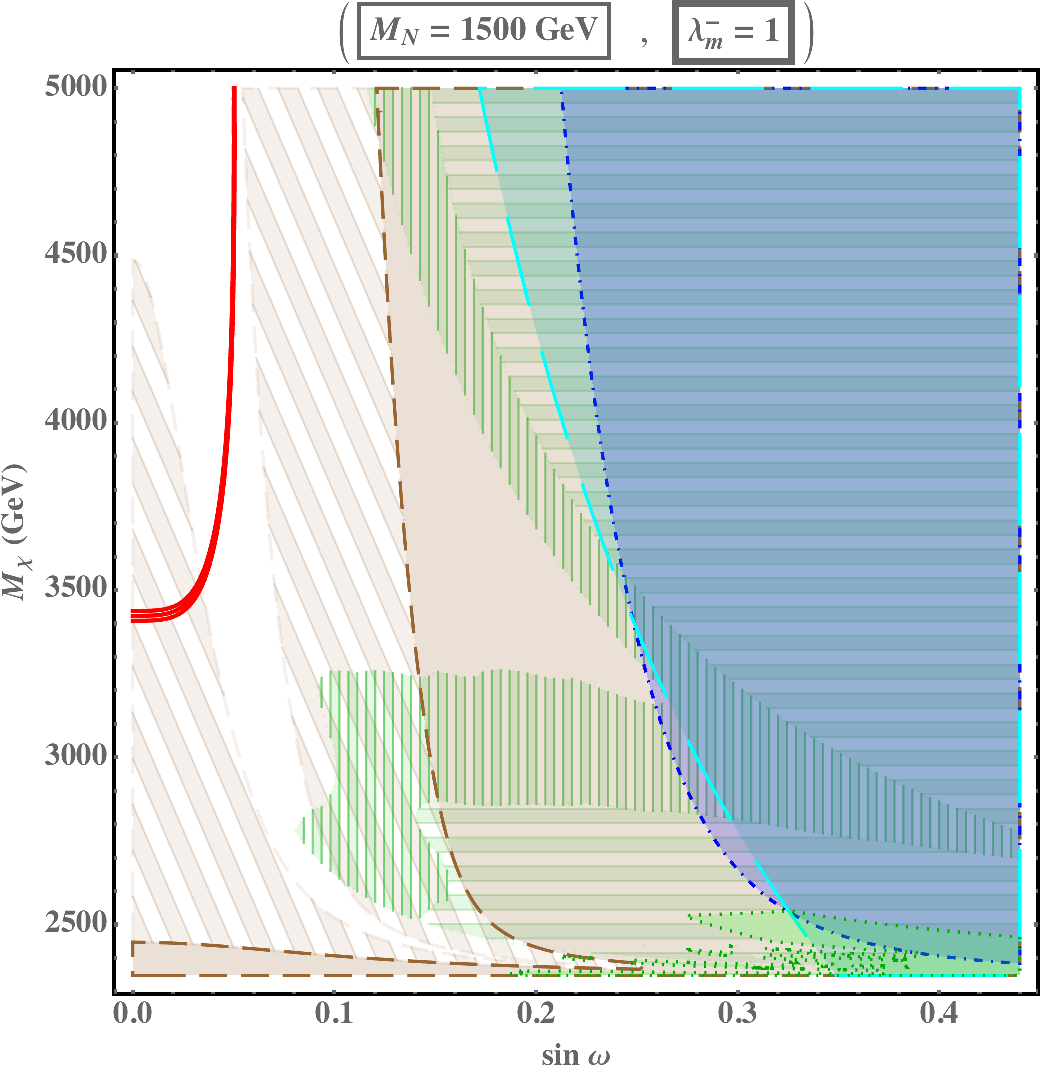}
\caption{Theoretical and experimental limits on the $\sin\omega-M_{\chi}$~plane, for benchmark values of $M_{N}$ and a larger $\lambda_{m}^{-} =1$. All solid-colored regions are excluded, whereas the striped regions correspond to the projected future bounds by the LHC and the XENON1T experiments. A large positive $\lambda_{m}^{-}$ allows for the potential XENON1T null results to be evaded, pushing the relic density curve into the viable region. (See the caption of Fig.~\ref{sinwms} for further details)}
\label{largerLmm}
\end{figure}

At this point, let us briefly comment on the possibility of having the TeV~scale $\chi$~pseudoscalar as the $\mathcal O (1)$~component of the dark matter, even facing a potential null result by the XENON1T experiment. Fig.~\ref{largerLmm} displays the $\sin\omega-M_{\chi}$~exclusion plots for two benchmark values of $M_{N}$ and a larger $\lambda_{m}^{-} =1$. It is evident from these plots that the constraints arising from a potential lack of discovery by the XENON1T experiment on the pseudoscalar dark matter relic density can be evaded for a sufficiently large and positive value of the $\lambda_{m}^{-}$ ~coupling, which pushes the relic abundance curve into the viable region of the parameter space. This, however, occurs at the expense of the cutoff not reaching the Planck scale, as discussed below.

Furthermore, we studied the vacuum stability and triviality conditions of the scenario and determined the viable region of the parameter space accommodating these conditions up to the Planck scale, extending the previous analysis with a cutoff at 100~TeV \cite{Farzinnia:2013pga}. We summarize, in various exclusion plots, the sensitivity of this viable region to the input parameters, combining the vacuum stability and triviality bounds with the prior theoretical and experimental constraints. We conclude that the mixing angle is, once more, restricted to small values $\sin\omega \lesssim 0.06$ by the triviality condition; whereas the dark matter mass is confined to several TeV by requiring the vacuum stability of the potential. In particular, accommodating the vacuum stability and triviality conditions up to the Planck scale requires $\lambda_{m}^{-} \sim +0.5$, given the sensitivity of the running of the other couplings to this parameter. For $\lambda_{m}^{-}$ lying near this particular value, a compromise between the stability and triviality of the couplings occurs, allowing for a (narrow) window of viability within the parameter space, where the cutoff may be extended to the Planck scale. This is in sharp contrast with the 100~TeV cutoff results, which could be accommodated by a variety of coupling values. Furthermore, increasing the cutoff to the Planck scals significantly reduces the allowed range of the mixing angle, in addition to confining the dark matter mass to the TeV ballpark. A TeV-scale right-handed Majorana neutrino mass, once more, inflicts substantial tension on the thermal relic abundance of the $\chi$~pseudoscalar.

As a final comment, we note that a small mixing angle, within the context of a classically scale invariant framework, immediately implies a large dynamically-generated singlet VEV, via the derived relation $\tan\omega = v_{\phi} / v_{\eta}$ (c.f. \eqref{hs}), which is a consequence of the scale symmetry. This, however, does not lead to a quadratic destabilization of the electroweak scale, since the magnitude of the mixing coupling becomes proportionally small (c.f. \eqref{lmp}). The dynamically-induced mass term for the Higgs doublet field \eqref{muSMfin}, giving rise to a successful spontaneous breaking of the electroweak symmetry, is therefore rather SM-like. We shall explore the consequences of a large singlet VEV, within the context of the current framework, in a forthcoming publication \cite{CFK}.

\section*{Acknowledgment}

We are grateful to Sekhar Chivukula for valuable comments on the manuscript. We also thank the anonymous referee for comments and clarifications regarding the additional right-handed Majorana neutrino operator. This work was supported by the IBS under the project code IBS-R018-D1.

\appendix*

\section{One-loop $\beta$-functions of the Theory}\label{betafun}

The one-loop $\beta$-functions for all the couplings of the current framework have already been computed in \cite{Farzinnia:2013pga}, and we summarize them here for the reference. The $\beta$-functions are defined as $d \mathcal C/d \log \mu = \beta_{\mathcal C}$, where $\mathcal C$ denotes any of the running couplings as a function of the renormalization scale~$\mu$. The SM gauge interactions are unaltered in the current scenario and their usual runnings apply (see e.g. \cite{Schrempp:1996fb})
\begin{equation} \label{SMgauge}
\pbrac{4\pi}^{2}\beta_{g} =-g^{3} \tbrac{+\frac{19}{6}} \ , \qquad \pbrac{4\pi}^{2}\beta_{g'} =+g'^{\,3} \tbrac{+\frac{41}{6}} \ , \qquad \pbrac{4\pi}^{2}\beta_{g_{c}} = -g_{c}^{3} \tbrac{+7} \ .
\end{equation}
The hypercharge coupling is normalized according to $g' = \sqrt{3/5} \, g_{1}$, with $g_{1}$ the corresponding coupling with the GUT normalization.

As the relevant Yukawa couplings, we consider those of the top quark and the flavor universal right-handed Majorana neutrinos. The latter are singlets under the SM gauge group and do not mix with the SM fermions. Therefore, we have
\begin{equation}\label{Yukawa}
\pbrac{4\pi}^{2}\beta_{y_{t}} =y_{t} \tbrac{ - \frac{9}{4} g^{2} - \frac{17}{12} g'^{\,2} -8 g_{c}^{2}+\frac{9}{2} y_{t}^{2} }\ , \qquad \pbrac{4\pi}^{2}\beta_{y_{N}} = 9 \, y_{N}^{3 } \ .
\end{equation}

Finally, with the normalizations employed in \eqref{V0}, the scalar quartic coupling $\beta$-functions, at one-loop, take the form
\begin{equation}\label{scalar}
\begin{split}
\pbrac{4\pi}^{2}\beta_{\lambda_{\phi}} =&\, +4\lambda_{\phi}^{2} + 3 \pbrac{\lambda_{m}^{+\, 2}+  \lambda_{m}^{-\, 2}} +3\lambda_{\phi} \tbrac{4 y_{t}^{2} - 3g^{2} - g'^{2}}- \frac{9}{4} \tbrac{16 y_{t}^{4} - 2 g^{4} - (g^{2} + g'^{2})^{2}} \ , \\
\pbrac{4\pi}^{2}\beta_{\lambda_{\eta}} =&\, +3\lambda_{\eta}^{2} + 12 \lambda_{m}^{+\, 2}+ 3 \lambda_{\eta\chi}^{2} +24\lambda_{\eta} y_{N}^{2}- 288 y_{N}^{4}  \ , \\
\pbrac{4\pi}^{2}\beta_{\lambda_{\chi}} =&\, +3\lambda_{\chi}^{2} + 12 \lambda_{m}^{-\, 2}+ 3 \lambda_{\eta\chi}^{2}  \ , \\
\pbrac{4\pi}^{2}\beta_{\lambda_{m}^{+}} =&\, +4\lambda_{m}^{+\, 2} + (2\lambda_{\phi} + \lambda_{\eta}) \lambda_{m}^{+} + \lambda_{\eta\chi} \lambda_{m}^{-} +\frac{3}{2}\lambda_{m}^{+} \tbrac{4 \pbrac{y_{t}^{2}+2y_{N}^{2}} - 3g^{2} - g'^{2}} \ ,\\
\pbrac{4\pi}^{2}\beta_{\lambda_{m}^{-}} =&\, +4\lambda_{m}^{-\, 2} + (2\lambda_{\phi} + \lambda_{\chi}) \lambda_{m}^{-} + \lambda_{\eta\chi} \lambda_{m}^{+} +\frac{3}{2}\lambda_{m}^{-} \tbrac{4 y_{t}^{2} - 3g^{2} - g'^{2}} \ ,\\
\pbrac{4\pi}^{2}\beta_{\lambda_{\eta\chi}} =&\, +4\lambda_{\eta\chi}^{2} + (\lambda_{\eta} + \lambda_{\chi}) \lambda_{\eta\chi} + 4\lambda_{m}^{+} \lambda_{m}^{-} +12\lambda_{\eta\chi}y_{N}^{2} \ .
\end{split}
\end{equation}


\begin{thebibliography}{99}


  
\bibitem{natural}
S.\ Weinberg, Phys.\ Lett.\ B\,{\bf 82} (1979) 387;
G.~'t Hooft,
NATO Adv.\ Study Inst.\ Ser.\ B Phys.\  {\bf 59}, 135 (1980).

\bibitem{smallcouple} 
R.~Foot, A.~Kobakhidze, K.~L.~McDonald and R.~R.~Volkas,
  Phys.\ Rev.\ D {\bf 77}, 035006 (2008)
  [arXiv:0709.2750 [hep-ph]];
  R.~Foot, A.~Kobakhidze, K.~L.~McDonald and R.~R.~Volkas,
  Phys.\ Rev.\ D {\bf 89}, no. 11, 115018 (2014)
  [arXiv:1310.0223 [hep-ph]];
  K.~Allison, C.~T.~Hill and G.~G.~Ross,
  Phys.\ Lett.\ B {\bf 738}, 191 (2014)
  [arXiv:1404.6268 [hep-ph]].
  

\bibitem{Bardeen}
W.\ A.\ Bardeen, ``On Naturalness in the Standard Model'',
FERMILAB-CONF-95-391-T, ``Beyond Higgs'', FERMILAB-CONF-08-118- T;
H.~Aoki and S.~Iso,
  Phys.\ Rev.\ D {\bf 86}, 013001 (2012)
  [arXiv:1201.0857 [hep-ph]].
  
\bibitem{Coleman:1973jx}
S.~R.~Coleman and E.~J.~Weinberg,
Phys.\ Rev.\ D {\bf 7} (1973) 1888.

\bibitem{LEPII} 
  R.~Barate {\it et al.}  [LEP Working Group for Higgs boson searches and ALEPH and DELPHI and L3 and OPAL Collaborations],
  Phys.\ Lett.\ B {\bf 565}, 61 (2003)
  [hep-ex/0306033].
For a summary of the results, see e.g.
  A.~Sopczak,
  Phys.\ Part.\ Nucl.\  {\bf 36}, 65 (2005)
  [Fiz.\ Elem.\ Chast.\ Atom.\ Yadra {\bf 36}, 127 (2005)]
  [hep-ph/0402231].
  
   \bibitem{Farzinnia:2013pga}
  A.~Farzinnia, H.~-J.~He and J.~Ren,
  Phys.\ Lett.\ B {\bf 727}, 141 (2013)
  [arXiv:1308.0295 [hep-ph]].
  
 \bibitem{LHCnew}
G.~Aad {\it et al.,}  [ATLAS Collaboration],
Phys.\ Lett.\ B\,{716} (2012) 1 [arXiv:1207.7214 [hep-ex]];
S.~Chatrchyan {\it et al.,}  [CMS Collaboration],
Phys.\ Lett.\ B\,{716} (2012) 30 [arXiv:1207.7235 [hep-ex]].

 \bibitem{SIother} 
 K.~A.~Meissner and H.~Nicolai,
  Phys.\ Lett.\ B {\bf 648}, 312 (2007)
  [hep-th/0612165];
   R.~Foot, A.~Kobakhidze, K.~L.~McDonald and R.~R.~Volkas,
  Phys.\ Rev.\ D {\bf 76}, 075014 (2007)
  [arXiv:0706.1829 [hep-ph]];
 K.~Ishiwata,
  Phys.\ Lett.\ B {\bf 710}, 134 (2012)
  [arXiv:1112.2696 [hep-ph]];
  D.~A.~Demir, M.~Frank and B.~Korutlu,
  Phys.\ Lett.\ B {\bf 728}, 393 (2014)
  [arXiv:1308.1203 [hep-ph]];
  T.~G.~Steele, Z.~W.~Wang, D.~Contreras and R.~B.~Mann,
  Phys.\ Rev.\ Lett.\  {\bf 112}, no. 17, 171602 (2014)
  [arXiv:1310.1960 [hep-ph]];
  M.~Lindner, S.~Schmidt and J.~Smirnov,
  JHEP {\bf 1410}, 177 (2014)
  [arXiv:1405.6204 [hep-ph]];
  Z.~Kang,
  arXiv:1411.2773 [hep-ph];
  K.~Kannike, G.~Hütsi, L.~Pizza, A.~Racioppi, M.~Raidal, A.~Salvio and A.~Strumia,
  JHEP {\bf 1505}, 065 (2015)
  [arXiv:1502.01334 [astro-ph.CO]].
  
\bibitem{Gabrielli:2013hma} 
  E.~Gabrielli, M.~Heikinheimo, K.~Kannike, A.~Racioppi, M.~Raidal and C.~Spethmann,
  Phys.\ Rev.\ D {\bf 89}, 015017 (2014)
  [arXiv:1309.6632 [hep-ph]].
  
\bibitem{AlexanderNunneley:2010nw} 
  L.~Alexander-Nunneley and A.~Pilaftsis,
  JHEP {\bf 1009}, 021 (2010)
  [arXiv:1006.5916 [hep-ph]].
 
    \bibitem{seesaw}
P.~Minkowski, Phys. Lett. B {\bf 67} (1977) 421;
T.~Yanagida, in \emph{Proceedings of the Workshop on the Unified
Theory and the Baryon Number in the Universe} (O.~Sawada and
A.~Sugamoto, eds.), KEK, Tsukuba, Japan, 1979, p.\,95;
M.~Gell-Mann, P.~Ramond, and R.~Slansky, \emph{Supergravity} (P.~van
Nieuwenhuizen et al., eds), North Holland, Amsterdam, 1979, p.\,315;
R.~N. Mohapatra and G.~Senjanovi{\'c}, Phys.\ Rev.\ Lett.\ {\bf 44} (1980) 912;
J.~Schechter and J.~W.~F.~Valle,
  Phys.\ Rev.\ D {\bf 22}, 2227 (1980).
  
  \bibitem{LHCHeavyH}
  [CMS Collaboration],
  CMS-PAS-HIG-13-002; CMS-PAS-HIG-13-003; 
  [ATLAS Collaboration],
  ATLAS-CONF-2013-013; ATLAS-CONF-2013-030.
  
  \bibitem{LUX2013} 
  D.~S.~Akerib {\it et al.}  [LUX Collaboration],
  arXiv:1310.8214 [astro-ph.CO].
  
\bibitem{Ade:2015xua} 
  P.~A.~R.~Ade {\it et al.} [Planck Collaboration],
  arXiv:1502.01589 [astro-ph.CO].
  
\bibitem{Farzinnia:2014xia} 
  A.~Farzinnia and J.~Ren,
  Phys.\ Rev.\ D {\bf 90}, no. 1, 015019 (2014)
  [arXiv:1405.0498 [hep-ph]].
  
\bibitem{Farzinnia:2014yqa} 
  A.~Farzinnia and J.~Ren,
  Phys.\ Rev.\ D {\bf 90}, no. 7, 075012 (2014)
  [arXiv:1408.3533 [hep-ph]].
  
     \bibitem{ATLASHeavyS}
  [ATLAS Collaboration],
 ATL-PHYS-PUB-2013-016.
 
 \bibitem{CMSHeavyS}
  [CMS Collaboration],
 CMS PAS FTR-13-024.
 
\bibitem{Aprile:2012zx} 
  E.~Aprile [XENON1T Collaboration],
  arXiv:1206.6288 [astro-ph.IM].
  
\bibitem{Gildener:1976ih}
E.~Gildener and S.~Weinberg,
Phys.\ Rev.\ D {\bf 13} (1976) 3333.
  
\bibitem{Cline:2013gha} 
  J.~M.~Cline, K.~Kainulainen, P.~Scott and C.~Weniger,
  Phys.\ Rev.\ D {\bf 88}, 055025 (2013)
  [arXiv:1306.4710 [hep-ph]].
  
\bibitem{Agrawal:2010fh} 
  P.~Agrawal, Z.~Chacko, C.~Kilic and R.~K.~Mishra,
  arXiv:1003.1912 [hep-ph].
  
\bibitem{Ellis:2000ds} 
  J.~R.~Ellis, A.~Ferstl and K.~A.~Olive,
  Phys.\ Lett.\ B {\bf 481}, 304 (2000)
  [hep-ph/0001005].
  
  \bibitem{Crivellin:2013ipa} 
  A.~Crivellin, M.~Hoferichter and M.~Procura,
  Phys.\ Rev.\ D {\bf 89}, 054021 (2014)
  [arXiv:1312.4951 [hep-ph]].
  
\bibitem{Zoller:2014cka} 
See M.~F.~Zoller, 
 arXiv:1411.2843 [hep-ph], and the references therein.
 
\bibitem{Degrassi:2012ry} 
  G.~Degrassi, S.~Di Vita, J.~Elias-Miro, J.~R.~Espinosa, G.~F.~Giudice, G.~Isidori and A.~Strumia,
  JHEP {\bf 1208}, 098 (2012)
  [arXiv:1205.6497 [hep-ph]].
  
\bibitem{Schrempp:1996fb} 
  B.~Schrempp and M.~Wimmer,
  Prog.\ Part.\ Nucl.\ Phys.\  {\bf 37}, 1 (1996)
  [hep-ph/9606386].
 
\bibitem{CFK} 
 K.~-Y.~Choi, A.~Farzinnia and S.~Kouwn,
 \textit{In preparation}.

 
  
\end{thebibliography}
\end{document}